
%
\catcode`@=11 
%
%
%
\font\seventeenrm=cmr10 scaled\magstep3
\font\fourteenrm=cmr10 scaled\magstep2
\font\twelverm=cmr10 scaled\magstep1
\font\ninerm=cmr9            \font\sixrm=cmr6

\font\fourteenbf=cmbx10 scaled\magstep2
\font\twelvebf=cmbx10 scaled\magstep1
\font\ninebf=cmbx9            \font\sixbf=cmbx6
\font\seventeeni=cmmi10 scaled\magstep3     \skewchar\seventeeni='177
\font\fourteeni=cmmi10 scaled\magstep2      \skewchar\fourteeni='177
\font\twelvei=cmmi10 scaled\magstep1        \skewchar\twelvei='177
\font\ninei=cmmi9                           \skewchar\ninei='177
\font\sixi=cmmi6                            \skewchar\sixi='177
\font\seventeensy=cmsy10 scaled\magstep3    \skewchar\seventeensy='60
\font\fourteensy=cmsy10 scaled\magstep2     \skewchar\fourteensy='60
\font\twelvesy=cmsy10 scaled\magstep1       \skewchar\twelvesy='60
\font\ninesy=cmsy9                          \skewchar\ninesy='60
\font\sixsy=cmsy6                           \skewchar\sixsy='60
%

%

\font\fourteensl=cmsl10 scaled\magstep2
\font\twelvesl=cmsl10 scaled\magstep1
\font\ninesl=cmsl9

\font\fourteenit=cmti10 scaled\magstep2
\font\twelveit=cmti10 scaled\magstep1
\font\twelvett=cmtt10 scaled\magstep1
\font\twelvecp=cmcsc10 scaled\magstep1
\font\tencp=cmcsc10
\newfam\cpfam
%
%
\newcount\f@ntkey            \f@ntkey=0
\def\samef@nt{\relax \ifcase\f@ntkey \rm \or\oldstyle \or\or
         \or\it \or\sl \or\bf \or\tt \or\caps \fi }
\def\fourteenpoint{\relax
    \textfont0=\fourteenrm          \scriptfont0=\tenrm
    \scriptscriptfont0=\sevenrm
     \def\rm{\fam0 \fourteenrm \f@ntkey=0 }\relax
    \textfont1=\fourteeni           \scriptfont1=\teni
    \scriptscriptfont1=\seveni
     \def\oldstyle{\fam1 \fourteeni\f@ntkey=1 }\relax
    \textfont2=\fourteensy          \scriptfont2=\tensy
    \scriptscriptfont2=\sevensy
    \def\it{\fam\itfam \fourteenit\f@ntkey=4 }\textfont\itfam=\fourteenit
    \def\sl{\fam\slfam \fourteensl\f@ntkey=5 }\textfont\slfam=\fourteensl
    \scriptfont\slfam=\tensl
    \def\bf{\fam\bffam \fourteenbf\f@ntkey=6 }\textfont\bffam=\fourteenbf
    \scriptfont\bffam=\tenbf     \scriptscriptfont\bffam=\sevenbf
    \def\tt{\fam\ttfam \twelvett \f@ntkey=7 }\textfont\ttfam=\twelvett
    \h@big=11.9\p@{} \h@Big=16.1\p@{} \h@bigg=20.3\p@{} \h@Bigg=24.5\p@{}
    \def\caps{\fam\cpfam \twelvecp \f@ntkey=8 }\textfont\cpfam=\twelvecp
    \setbox\strutbox=\hbox{\vrule height 12pt depth 5pt width\z@}
    \samef@nt}
\def\twelvepoint{\relax
    \textfont0=\twelverm          \scriptfont0=\ninerm
    \scriptscriptfont0=\sixrm
     \def\rm{\fam0 \twelverm \f@ntkey=0 }\relax
    \textfont1=\twelvei           \scriptfont1=\ninei
    \scriptscriptfont1=\sixi
     \def\oldstyle{\fam1 \twelvei\f@ntkey=1 }\relax
    \textfont2=\twelvesy          \scriptfont2=\ninesy
    \scriptscriptfont2=\sixsy
    \def\it{\fam\itfam \twelveit \f@ntkey=4 }\textfont\itfam=\twelveit
    \def\sl{\fam\slfam \twelvesl \f@ntkey=5 }\textfont\slfam=\twelvesl
    \scriptfont\slfam=\ninesl
    \def\bf{\fam\bffam \twelvebf \f@ntkey=6 }\textfont\bffam=\twelvebf
    \scriptfont\bffam=\ninebf     \scriptscriptfont\bffam=\sixbf
    \def\tt{\fam\ttfam \twelvett \f@ntkey=7 }\textfont\ttfam=\twelvett
    \h@big=10.2\p@{}
    \h@Big=13.8\p@{}
    \h@bigg=17.4\p@{}
    \h@Bigg=21.0\p@{}
    \def\caps{\fam\cpfam \twelvecp \f@ntkey=8 }\textfont\cpfam=\twelvecp
    \setbox\strutbox=\hbox{\vrule height 10pt depth 4pt width\z@}
    \samef@nt}
\def\tenpoint{\relax
    \textfont0=\tenrm          \scriptfont0=\sevenrm
    \scriptscriptfont0=\fiverm
    \def\rm{\fam0 \tenrm \f@ntkey=0 }\relax
    \textfont1=\teni           \scriptfont1=\seveni
    \scriptscriptfont1=\fivei
    \def\oldstyle{\fam1 \teni \f@ntkey=1 }\relax
    \textfont2=\tensy          \scriptfont2=\sevensy
    \scriptscriptfont2=\fivesy
    \def\it{\fam\itfam \tenit \f@ntkey=4 }\textfont\itfam=\tenit
    \def\sl{\fam\slfam \tensl \f@ntkey=5 }\textfont\slfam=\tensl
    \def\bf{\fam\bffam \tenbf \f@ntkey=6 }\textfont\bffam=\tenbf
    \scriptfont\bffam=\sevenbf     \scriptscriptfont\bffam=\fivebf
    \def\tt{\fam\ttfam \tentt \f@ntkey=7 }\textfont\ttfam=\tentt
    \def\caps{\fam\cpfam \tencp \f@ntkey=8 }\textfont\cpfam=\tencp
    \setbox\strutbox=\hbox{\vrule height 8.5pt depth 3.5pt width\z@}
    \samef@nt}
%
%
%
%
\newdimen\h@big  \h@big=8.5\p@
\newdimen\h@Big  \h@Big=11.5\p@
\newdimen\h@bigg  \h@bigg=14.5\p@
\newdimen\h@Bigg  \h@Bigg=17.5\p@
\def\big#1{{\hbox{$\left#1\vbox to\h@big{}\right.\n@space$}}}
\def\Big#1{{\hbox{$\left#1\vbox to\h@Big{}\right.\n@space$}}}
\def\bigg#1{{\hbox{$\left#1\vbox to\h@bigg{}\right.\n@space$}}}
\def\Bigg#1{{\hbox{$\left#1\vbox to\h@Bigg{}\right.\n@space$}}}
%
%
%
\normalbaselineskip = 20pt plus 0.2pt minus 0.1pt
\normallineskip = 1.5pt plus 0.1pt minus 0.1pt
\normallineskiplimit = 1.5pt
\newskip\normaldisplayskip
\normaldisplayskip = 18pt plus 5pt minus 10pt
\newskip\normaldispshortskip
\normaldispshortskip = 6pt plus 5pt
\newskip\normalparskip
\normalparskip = 6pt plus 2pt minus 1pt
\newskip\skipregister
\skipregister = 5pt plus 2pt minus 1.5pt
\newif\ifsingl@    \newif\ifdoubl@
\newif\iftwelv@    \twelv@true
\def\singlespace{\singl@true\doubl@false\spaces@t}
\def\doublespace{\singl@false\doubl@true\spaces@t}
\def\normalspace{\singl@false\doubl@false\spaces@t}
\def\Tenpoint{\tenpoint\twelv@false\spaces@t}
\def\Twelvepoint{\twelvepoint\twelv@true\spaces@t}
\def\spaces@t{\relax%
 \iftwelv@ \ifsingl@\subspaces@t3:4;\else\subspaces@t1:1;\fi%
 \else \ifsingl@\subspaces@t3:5;\else\subspaces@t4:5;\fi \fi%
 \ifdoubl@ \multiply\baselineskip by 5%
 \divide\baselineskip by 4 \fi \unskip}
\def\subspaces@t#1:#2;{
      \baselineskip = \normalbaselineskip
      \multiply\baselineskip by #1 \divide\baselineskip by #2
      \lineskip = \normallineskip
      \multiply\lineskip by #1 \divide\lineskip by #2
      \lineskiplimit = \normallineskiplimit
      \multiply\lineskiplimit by #1 \divide\lineskiplimit by #2
      \parskip = \normalparskip
      \multiply\parskip by #1 \divide\parskip by #2
      \abovedisplayskip = \normaldisplayskip
      \multiply\abovedisplayskip by #1 \divide\abovedisplayskip by #2
      \belowdisplayskip = \abovedisplayskip
      \abovedisplayshortskip = \normaldispshortskip
      \multiply\abovedisplayshortskip by #1
        \divide\abovedisplayshortskip by #2
      \belowdisplayshortskip = \abovedisplayshortskip
      \advance\belowdisplayshortskip by \belowdisplayskip
      \divide\belowdisplayshortskip by 2
      \smallskipamount = \skipregister
      \multiply\smallskipamount by #1 \divide\smallskipamount by #2
      \medskipamount = \smallskipamount \multiply\medskipamount by 2
      \bigskipamount = \smallskipamount \multiply\bigskipamount by 4 }
\def\normalbaselines{ \baselineskip=\normalbaselineskip
   \lineskip=\normallineskip \lineskiplimit=\normallineskip
   \iftwelv@\else \multiply\baselineskip by 4 \divide\baselineskip by 5
     \multiply\lineskiplimit by 4 \divide\lineskiplimit by 5
     \multiply\lineskip by 4 \divide\lineskip by 5 \fi }
\Twelvepoint  
\interlinepenalty=50
\interfootnotelinepenalty=5000
\predisplaypenalty=9000
\postdisplaypenalty=500
\hfuzz=1pt
\vfuzz=0.2pt
%
%
%
\def\pagecontents{
   \ifvoid\topins\else\unvbox\topins\vskip\skip\topins\fi
   \dimen@ = \dp255 \unvbox255
   \ifvoid\footins\else\vskip\skip\footins\footrule\unvbox\footins\fi
   \ifr@ggedbottom \kern-\dimen@ \vfil \fi }
\def\makeheadline{\vbox to 0pt{ \skip@=\topskip
      \advance\skip@ by -12pt \advance\skip@ by -2\normalbaselineskip
      \vskip\skip@ \line{\vbox to 12pt{}\the\headline} \vss
      }\nointerlineskip}
\def\makefootline{\baselineskip = 1.5\normalbaselineskip
                 \line{\the\footline}}
\newif\iffrontpage
\newif\ifletterstyle
\newif\ifp@genum
\def\nopagenumbers{\p@genumfalse}
\def\pagenumbers{\p@genumtrue}
\pagenumbers
\newtoks\paperheadline
\newtoks\letterheadline
\newtoks\letterfrontheadline
\newtoks\lettermainheadline
\newtoks\paperfootline
\newtoks\letterfootline
\newtoks\date
\footline={\ifletterstyle\the\letterfootline\else\the\paperfootline\fi}
\paperfootline={\hss\iffrontpage\else\ifp@genum\tenrm\folio\hss\fi\fi}
\letterfootline={\hfil}
\headline={\ifletterstyle\the\letterheadline\else\the\paperheadline\fi}
\paperheadline={\hfil}
\letterheadline{\iffrontpage\the\letterfrontheadline
     \else\the\lettermainheadline\fi}
\lettermainheadline={\rm\ifp@genum page \ \folio\fi\hfil\the\date}
\def\monthname{\relax\ifcase\month 0/\or January\or February\or
   March\or April\or May\or June\or July\or August\or September\or
   October\or November\or December\else\number\month/\fi}
\date={\monthname\ \number\day, \number\year}
\countdef\pagenumber=1  \pagenumber=1
\def\advancepageno{\global\advance\pageno by 1
   \ifnum\pagenumber<0 \global\advance\pagenumber by -1
    \else\global\advance\pagenumber by 1 \fi \global\frontpagefalse }
\def\folio{\ifnum\pagenumber<0 \romannumeral-\pagenumber
           \else \number\pagenumber \fi }
\def\footrule{\dimen@=\prevdepth\nointerlineskip
   \vbox to 0pt{\vskip -0.25\baselineskip \hrule width 0.35\hsize \vss}
   \prevdepth=\dimen@ }
\newtoks\foottokens
\foottokens={\Tenpoint\singlespace}
\newdimen\footindent
\footindent=24pt
\def\vfootnote#1{\insert\footins\bgroup  \the\foottokens
   \interlinepenalty=\interfootnotelinepenalty \floatingpenalty=20000
   \splittopskip=\ht\strutbox \boxmaxdepth=\dp\strutbox
   \leftskip=\footindent \rightskip=\z@skip
   \parindent=0.5\footindent \parfillskip=0pt plus 1fil
   \spaceskip=\z@skip \xspaceskip=\z@skip
   \Textindent{$ #1 $}\footstrut\futurelet\next\fo@t}
\def\Textindent#1{\noindent\llap{#1\enspace}\ignorespaces}
\def\footnote#1{\attach{#1}\vfootnote{#1}}

\def\foot{\attach\footsymbolgen\vfootnote{\footsymbol}}
\let\footsymbol=\star
\newcount\lastf@@t           \lastf@@t=-1
\newcount\footsymbolcount    \footsymbolcount=0
\newif\ifPhysRev
\def\footsymbolgen{\relax \ifPhysRev \iffrontpage \NPsymbolgen\else
      \PRsymbolgen\fi \else \NPsymbolgen\fi
   \global\lastf@@t=\pageno \footsymbol }
\def\NPsymbolgen{\ifnum\footsymbolcount<0 \global\footsymbolcount=0\fi
   {\iffrontpage \else \advance\lastf@@t by 1 \fi
    \ifnum\lastf@@t<\pageno \global\footsymbolcount=0
     \else \global\advance\footsymbolcount by 1 \fi }
   \ifcase\footsymbolcount \fd@f\star\or \fd@f\diamond\or \fd@f\ast\or
\fd@f\dagger\or \fd@f\ddagger\or \fd@f\natural\or \fd@f\bullet\or
    \fd@f\nabla\else \fd@f\dagger\global\footsymbolcount=0 \fi }
\def\fd@f#1{\xdef\footsymbol{#1}}
\def\PRsymbolgen{\ifnum\footsymbolcount>0 \global\footsymbolcount=0\fi
      \global\advance\footsymbolcount by -1
      \xdef\footsymbol{\sharp\number-\footsymbolcount} }
\def\space@ver#1{\let\@sf=\empty \ifmmode #1\else \ifhmode
   \edef\@sf{\spacefactor=\the\spacefactor}\unskip${}#1$\relax\fi\fi}
\def\attach#1{\space@ver{\strut^{\mkern 2mu #1} }\@sf\ }
%
%
%
\newcount\chapternumber      \chapternumber=0
\newcount\sectionnumber      \sectionnumber=0
\newcount\equanumber         \equanumber=0
\let\chapterlabel=0
\newtoks\chapterstyle        \chapterstyle={\Number}
\newskip\chapterskip         \chapterskip=\bigskipamount
\newskip\sectionskip         \sectionskip=\medskipamount
\newskip\headskip            \headskip=8pt plus 3pt minus 3pt
\newdimen\chapterminspace    \chapterminspace=15pc
\newdimen\sectionminspace    \sectionminspace=10pc
\newdimen\referenceminspace  \referenceminspace=25pc
\def\chapterreset{\global\advance\chapternumber by 1
   \ifnum\the\equanumber<0 \else\global\equanumber=0\fi
   \sectionnumber=0 \makel@bel}
\def\makel@bel{\xdef\chapterlabel{%
\the\chapterstyle{\the\chapternumber}.}}
\def\sectionlabel{{\number\sectionnumber}. \quad }
\def\alphabetic#1{\count255='140 \advance\count255 by #1\char\count255}
\def\Alphabetic#1{\count255='100 \advance\count255 by #1\char\count255}
\def\Roman#1{\uppercase\expandafter{\romannumeral #1}}
\def\roman#1{\romannumeral #1}
\def\Number#1{\number #1}
\def\unnumberedchapters{\let\makel@bel=\relax \let\chapterlabel=\relax
\let\sectionlabel=\relax \equanumber=-1 }
\def\titlestyle#1{\par\begingroup \interlinepenalty=9999
     \leftskip=0.02\hsize plus 0.23\hsize minus 0.02\hsize
     \rightskip=\leftskip \parfillskip=0pt
     \hyphenpenalty=9000 \exhyphenpenalty=9000
     \tolerance=9999 \pretolerance=9000
     \spaceskip=0.333em \xspaceskip=0.5em
     \iftwelv@\fourteenpoint\else\twelvepoint\fi
   \noindent #1\par\endgroup }
\def\spacecheck#1{\dimen@=\pagegoal\advance\dimen@ by -\pagetotal
   \ifdim\dimen@<#1 \ifdim\dimen@>0pt \vfil\break \fi\fi}
\def\chapter#1{\par \penalty-300 \vskip\chapterskip
   \spacecheck\chapterminspace
   \chapterreset \titlestyle{\chapterlabel \ #1}
   \nobreak\vskip\headskip \penalty 30000
   \wlog{\string\chapter\ \chapterlabel} }

\def\section#1{\par \ifnum\the\lastpenalty=30000\else
   \penalty-200\vskip\sectionskip \spacecheck\sectionminspace\fi
   \wlog{\string\section\ \chapterlabel \the\sectionnumber}
   \global\advance\sectionnumber by 1  \noindent
   {\sl \enspace \chapterlabel \sectionlabel #1}\par
   \nobreak\vskip\headskip \penalty 30000 }
\def\subsection#1{\par
   \ifnum\the\lastpenalty=30000\else \penalty-100\smallskip \fi
   \noindent\undertext{#1}\enspace \vadjust{\penalty5000}}

\def\undertext#1{\vtop{\hbox{#1}\kern 1pt \hrule}}
\def\ack{\par\penalty-100\medskip \spacecheck\sectionminspace
   \line{\fourteenrm\hfil Acknowledgements\hfil}\nobreak\vskip\headskip }
\def\APPENDIX#1#2{\par\penalty-300\vskip\chapterskip
   \spacecheck\chapterminspace \chapterreset \xdef\chapterlabel{#1}
   \titlestyle{Appendix #2} \nobreak\vskip\headskip \penalty 30000
   \wlog{\string\Appendix\ \chapterlabel} }
\def\Appendix#1{\APPENDIX{#1}{#1}}
\def\appendix{\APPENDIX{A}{}}
%
%
%
\def\eqname#1{\relax \ifnum\the\equanumber<0
     \xdef#1{{\rm(\number-\equanumber)}}\global\advance\equanumber by -1
    \else \global\advance\equanumber by 1
      \xdef#1{{\rm(\chapterlabel \number\equanumber)}} \fi}
\def\eq{\eqname\?\?}

\def\eqinsert#1{\noalign{\dimen@=\prevdepth \nointerlineskip
   \setbox0=\hbox to\displaywidth{\hfil #1}
   \vbox to 0pt{\vss\hbox{$\!\box0\!$}\kern-0.5\baselineskip}
   \prevdepth=\dimen@}}
%

%

%

%
%
\def\GENITEM#1;#2{\par \hangafter=0 \hangindent=#1
    \Textindent{$ #2 $}\ignorespaces}
\outer\def\newitem#1=#2;{\gdef#1{\GENITEM #2;}}
\newdimen\itemsize                \itemsize=30pt
\newitem\item=1\itemsize;
\newitem\sitem=1.75\itemsize;     
\newitem\ssitem=2.5\itemsize;     
\outer\def\newlist#1=#2&#3&#4;{\toks0={#2}\toks1={#3}%
   \count255=\escapechar \escapechar=-1
   \alloc@0\list\countdef\insc@unt\listcount     \listcount=0
   \edef#1{\par
      \countdef\listcount=\the\allocationnumber
      \advance\listcount by 1
      \hangafter=0 \hangindent=#4
      \Textindent{\the\toks0{\listcount}\the\toks1}}
   \expandafter\expandafter\expandafter
    \edef\c@t#1{begin}{\par
      \countdef\listcount=\the\allocationnumber \listcount=1
      \hangafter=0 \hangindent=#4
      \Textindent{\the\toks0{\listcount}\the\toks1}}
   \expandafter\expandafter\expandafter
    \edef\c@t#1{con}{\par \hangafter=0 \hangindent=#4 \noindent}
   \escapechar=\count255}
\def\c@t#1#2{\csname\string#1#2\endcsname}
\newlist\point=\Number&.&1.0\itemsize;
\newlist\subpoint=(\alphabetic&)&1.75\itemsize;
\newlist\subsubpoint=(\roman&)&2.5\itemsize;
%

%
%
%
\newcount\referencecount     \referencecount=0
\newif\ifreferenceopen       \newwrite\referencewrite
\newtoks\rw@toks
\def\NPrefmark#1{\attach{\scriptscriptstyle [ #1 ] }}
\let\PRrefmark=\attach
\def\refmark#1{\relax\ifPhysRev\PRrefmark{#1}\else\NPrefmark{#1}\fi}
\def\refend{\refmark{\number\referencecount}}
\newcount\lastrefsbegincount \lastrefsbegincount=0
\def\refsend{\refmark{\count255=\referencecount
   \advance\count255 by-\lastrefsbegincount
   \ifcase\count255 \number\referencecount
   \or \number\lastrefsbegincount,\number\referencecount
   \else \number\lastrefsbegincount-\number\referencecount \fi}}
\def\refch@ck{\chardef\rw@write=\referencewrite
   \ifreferenceopen \else \referenceopentrue
   \immediate\openout\referencewrite=referenc.txa \fi}
%
{\catcode`\^^M=\active 
  \gdef\obeyendofline{\catcode`\^^M\active \let^^M\ }}%
%
{\catcode`\^^M=\active 
  \gdef\ignoreendofline{\catcode`\^^M=5}}
{\obeyendofline\gdef\rw@start#1{\def\t@st{#1} \ifx\t@st\blankend%
\endgroup \@sf \relax \else \ifx\t@st\bl@nkend \endgroup \@sf \relax%
\else \rw@begin#1
\backtotext
\fi \fi } }
{\obeyendofline\gdef\rw@begin#1
{\def\n@xt{#1}\rw@toks={#1}\relax%
\rw@next}}
\def\blankend{}
{\obeylines\gdef\bl@nkend{
}}
\newif\iffirstrefline  \firstreflinetrue
\def\rwr@teswitch{\ifx\n@xt\blankend \let\n@xt=\rw@begin %
 \else\iffirstrefline \global\firstreflinefalse%
\immediate\write\rw@write{\noexpand\obeyendofline \the\rw@toks}%
\let\n@xt=\rw@begin%
      \else\ifx\n@xt\rw@@d \def\n@xt{\immediate\write\rw@write{%
        \noexpand\ignoreendofline}\endgroup \@sf}%
             \else \immediate\write\rw@write{\the\rw@toks}%
             \let\n@xt=\rw@begin\fi\fi \fi}
\def\rw@next{\rwr@teswitch\n@xt}
\def\rw@@d{\backtotext} \let\rw@end=\relax
\let\backtotext=\relax

\newdimen\refindent     \refindent=30pt
\def\refitem#1{\par \hangafter=0 \hangindent=\refindent \Textindent{#1}}
\def\REFNUM#1{\space@ver{}\refch@ck \firstreflinetrue%
 \global\advance\referencecount by 1 \xdef#1{\the\referencecount}}
\def\refnum#1{\space@ver{}\refch@ck \firstreflinetrue%
 \global\advance\referencecount by 1 \xdef#1{\the\referencecount}\refend}

\def\REF#1{\REFNUM#1%
 \immediate\write\referencewrite{%
 \noexpand\refitem{#1.}}%
\begingroup\obeyendofline\rw@start}
\def\ref{\refnum\?%
 \immediate\write\referencewrite{\noexpand\refitem{\?.}}%
\begingroup\obeyendofline\rw@start}
\def\Ref#1{\refnum#1%
 \immediate\write\referencewrite{\noexpand\refitem{#1.}}%
\begingroup\obeyendofline\rw@start}
\def\REFS#1{\REFNUM#1\global\lastrefsbegincount=\referencecount
\immediate\write\referencewrite{\noexpand\refitem{#1.}}%
\begingroup\obeyendofline\rw@start}
\newcount\figurecount     \figurecount=0
\newif\iffigureopen       \newwrite\figurewrite
\def\figch@ck{\chardef\rw@write=\figurewrite \iffigureopen\else
   \immediate\openout\figurewrite=figures.txa
   \figureopentrue\fi}
\def\FIGNUM#1{\space@ver{}\figch@ck \firstreflinetrue%
 \global\advance\figurecount by 1 \xdef#1{\the\figurecount}}
\def\FIG#1{\FIGNUM#1
   \immediate\write\figurewrite{\noexpand\refitem{#1.}}%
   \begingroup\obeyendofline\rw@start}
\def\fig{\FIGNUM\? fig.^^\?%
\immediate\write\figurewrite{\noexpand\refitem{\?.}}%
\begingroup\obeyendofline\rw@start}
\def\figure{\FIGNUM\? figure^^\?
   \immediate\write\figurewrite{\noexpand\refitem{\?.}}%
   \begingroup\obeyendofline\rw@start}
\def\Fig{\FIGNUM\? Fig.^^\?%
\immediate\write\figurewrite{\noexpand\refitem{\?.}}%
\begingroup\obeyendofline\rw@start}
\def\Figure{\FIGNUM\? Figure^^\?%
\immediate\write\figurewrite{\noexpand\refitem{\?.}}%
\begingroup\obeyendofline\rw@start}
\newcount\tablecount     \tablecount=0
\newif\iftableopen       \newwrite\tablewrite
\def\tabch@ck{\chardef\rw@write=\tablewrite \iftableopen\else
   \immediate\openout\tablewrite=tables.txa
   \tableopentrue\fi}
\def\TABNUM#1{\space@ver{}\tabch@ck \firstreflinetrue%
 \global\advance\tablecount by 1 \xdef#1{\the\tablecount}}
\def\TABLE#1{\TABNUM#1
   \immediate\write\tablewrite{\noexpand\refitem{#1.}}%
   \begingroup\obeyendofline\rw@start}
\def\Table{\TABNUM\? Table^^\?%
\immediate\write\tablewrite{\noexpand\refitem{\?.}}%
\begingroup\obeyendofline\rw@start}
%

%
%
%
\def\masterreset{\global\pagenumber=1 \global\chapternumber=0
   \ifnum\the\equanumber<0\else \global\equanumber=0\fi
   \global\sectionnumber=0
   \global\referencecount=0 \global\figurecount=0 \global\tablecount=0 }
\def\FRONTPAGE{\ifvoid255\else\vfill\penalty-2000\fi
      \masterreset\global\frontpagetrue
      \global\lastf@@t=0 \global\footsymbolcount=0}

\def\paperstyle{\letterstylefalse\normalspace\papersize}
\def\letterstyle{\letterstyletrue\singlespace\lettersize}
\def\papersize{\hsize=35pc\vsize=52pc\hoffset=0.5truecm\voffset=-2pc
               \skip\footins=\bigskipamount}
\def\lettersize{\hsize=6.5in\vsize=8.5in\hoffset=0in\voffset=1in
   \skip\footins=\smallskipamount \multiply\skip\footins by 3 }
\paperstyle   
%
%
\def\MEMO{\letterstyle\FRONTPAGE \letterfrontheadline={\hfil}
    \line{\quad\fourteenrm EFI MEMORANDUM\hfil\twelverm\the\date\quad}
    \medskip \memod@f}

\def\memit@m#1{\smallskip \hangafter=0 \hangindent=1in
      \Textindent{\caps #1}}
\def\memod@f{\xdef\to{\memit@m{To:}}\xdef\from{\memit@m{From:}}%
     \xdef\topic{\memit@m{Topic:}}\xdef\subject{\memit@m{Subject:}}%
     \xdef\rule{\bigskip\hrule height 1pt\bigskip}}
\memod@f
\newskip\lettertopfil
\lettertopfil = 0pt plus 1.5in minus 0pt
\newskip\letterbottomfil
\letterbottomfil = 0pt plus 2.3in minus 0pt
\newskip\spskip \setbox0\hbox{\ } \spskip=-1\wd0

\newskip\signatureskip       \signatureskip=40pt
\def\signed#1{\vskip 0.7cm minus .5cm\par \penalty 9000 \bigskip \dt@pfalse
  \everycr={\noalign{\ifdt@p\vskip\signatureskip\global\dt@pfalse\fi}}
  \setbox0=\vbox{\singlespace \halign{\tabskip 0pt \strut ##\hfil\cr
   \noalign{\global\dt@ptrue}#1\crcr}}
  \line{\hskip 0.5\hsize minus 0.5\hsize \box0\hfil} \medskip }

\def\endletter{\ifnum\pagenumber=1 \vskip\letterbottomfil\supereject
\else \vfil\supereject \fi}
\newbox\letterb@x
\def\lettertext{\par\unvcopy\letterb@x\par}
\def\multiletter{\setbox\letterb@x=\vbox\bgroup
      \everypar{\vrule height 1\baselineskip depth 0pt width 0pt }
      \singlespace \topskip=\baselineskip }
\def\letterend{\par\egroup}
%
%
%
\newskip\frontpageskip
\newtoks\pubtype
\newtoks\bisasPubnum
\newtoks\icPubnum
\newtoks\bisaspubnum
\newtoks\icpubnum
\newif\ifp@bblock  \p@bblocktrue
\def\PH@SR@V{\doubl@true \baselineskip=24.1pt plus 0.2pt minus 0.1pt
             \parskip= 3pt plus 2pt minus 1pt }
\def\PHYSREV{\paperstyle\PhysRevtrue\PH@SR@V}
\def\titlepage{\FRONTPAGE\paperstyle\ifPhysRev\PH@SR@V\fi
   \ifp@bblock\p@bblock\fi}
\def\nopubblock{\p@bblockfalse}
\def\endpage{\vfil\break}
\frontpageskip=1\medskipamount plus .5fil
\pubtype={\tensl Preliminary Version}
\bisasPubnum={$\caps POLFIS-TH18-91\the\bisaspubnum $}
\icPubnum={$\caps IC/\the\icpubnum $}
\bisaspubnum={0000}
\icpubnum={0000}
\date={\monthname, \number\year}
\def\p@bblock{\begingroup \tabskip=\hsize minus \hsize
   \baselineskip=1.5\ht\strutbox \topspace-2\baselineskip
   \halign to\hsize{\strut ##\hfil\tabskip=0pt\crcr
   \the\bisasPubnum\cr
   \the\date\cr  \the\pubtype\cr
   }\endgroup}
\def\title#1{\vskip\frontpageskip \titlestyle{\seventeenrm #1}\vskip\headskip}
\def\author#1{\vskip\frontpageskip\titlestyle{\fourteenrm #1}\nobreak}
\def\andauthor{\vskip\frontpageskip\centerline{and}\author}

\def\address#1{\par\titlestyle{\twelvepoint\it #1}}

\def\abstract{\vskip\smallskipamount\vskip\frontpageskip
              \centerline{\fourteenrm ABSTRACT}
              \vskip\headskip }

%
%
%
\def\ie{\hbox{\it i.e.}}     
\def\eg{\hbox{\it e.g.}}     

\def\\{\relax\ifmmode\backslash\else$\backslash$\fi}
\def\globaleqnumbers{\relax\ifnum\the\equanumber<0%
\else\global\equanumber=-1\fi}
\def\nextline{\unskip\nobreak\hskip\parfillskip\break}

\def\journal#1&#2(#3){\unskip, \sl #1\ \bf #2 \rm(19#3) }

\def\topspace{\hrule height 0pt depth 0pt \vskip}

\let\int=\intop         
\def\prop{\mathrel{{\mathchoice{\pr@p\scriptstyle}{\pr@p\scriptstyle}{
                \pr@p\scriptscriptstyle}{\pr@p\scriptscriptstyle} }}}
\def\pr@p#1{\setbox0=\hbox{$\cal #1 \char'103$}
   \hbox{$\cal #1 \char'117$\kern-.4\wd0\box0}}
\def\lsim{\mathrel{\mathpalette\@versim<}}
\def\gsim{\mathrel{\mathpalette\@versim>}}
\def\@versim#1#2{\lower0.2ex\vbox{\baselineskip\z@skip\lineskip\z@skip
  \lineskiplimit\z@\ialign{$\m@th#1\hfil##\hfil$\crcr#2\crcr\sim\crcr}}}
%
%
%
\let\sec@nt=\sec
\def\sec{\relax\ifmmode\let\n@xt=\sec@nt\else\let\n@xt\section\fi\n@xt}
\def\obsolete#1{\message{Macro \string #1 is obsolete.}}
\def\firstsec#1{\obsolete\firstsec \section{#1}}
\def\firstsubsec#1{\obsolete\firstsubsec \subsection{#1}}
\def\thispage#1{\obsolete\thispage \global\pagenumber=#1\frontpagefalse}
\def\thischapter#1{\obsolete\thischapter \global\chapternumber=#1}
\def\nextequation#1{\obsolete\nextequation \global\equanumber=#1
   \ifnum\the\equanumber>0 \global\advance\equanumber by 1 \fi}
\def\BOXITEM{\afterassigment\B@XITEM\setbox0=}
\def\B@XITEM{\par\hangindent\wd0 \noindent\box0 }
%

%
\catcode`@=12 
%
\everyjob{\input myphyx }
\voffset 2.truecm
\def\frac#1#2{{#1\over#2}}

\def\andjournal#1&#2(#3){\sl #1^^\bf #2 \rm (19#3) }

\catcode`\@=11
\def\slash#1{\mathord{\mathpalette\c@ncel{#1}}}
\overfullrule=0pt
\def\marginnote#1{}
\def\steepslash{\c@ncel}

\def\frac#1#2{{#1\over #2}}

\def\footnote{\foot}

\def\section#1{\par \ifnum\the\lastpenalty=30000\else
   \penalty-200\vskip\sectionskip \spacecheck\sectionminspace\fi
   \wlog{\string\section\ \the\sectionnumber}
   \global\advance\sectionnumber by 1  \noindent
   {\caps\enspace\sectionlabel #1}\par
   \nobreak\vskip\headskip \penalty 30000 }
\catcode`\@=12
\hyphenation{su-per-sur-face}
\hyphenation{su-per-sur-fa-ces}
\hyphenation{-su-per-dif-fer-en-tial}
\hyphenation{-su-per-dif-fer-en-tials}
\pubtype={}
\bisaspubnum={ }
\icpubnum={}

\date={December,1991}

\def\zet{{Z \kern-.45em Z}}
\def\complex{{\kern .1em {\raise .47ex \hbox {$\scriptscriptstyle |$}}
\kern -.4em {\rm C}}}
\def\real{{\vrule height 1.6ex width 0.05em depth 0ex
\kern -0.06em {\rm R}}}
\def\rational{{\kern .1em {\raise .47ex \hbox{$\scripscriptstyle |$}}
\kern -.35em {\rm Q}}}

\def\dbar{{\bar D}}
\def\det{{\rm Det}}
\def\lapbar{{\bar\triangle}}

\def\CcC{{\hbox{\tenrm C\kern-.45em{\vrule height.67em width0.08em depth-.04em
\hskip.45em }}}}
\def\RrR{{\hbox{\tenrm I\kern-.17em{R}}}}
\def\HhH{{\hbox{\tenrm {I\kern-.18em{H}}\kern-.18em{I}}}}
\def\DdD{{\hbox{\tenrm {I\kern-.18em{D}}\kern-.36em {\vrule height.62em
width0.08em depth-.04em\hskip.36em}}}}
\def\ZzZ{{\hbox{\tenrm Z\kern-.31em{Z}}}}
\def\IiI{{\hbox{\tenrm I\kern-.19em{I}}}}
\def\NnN{{\hbox{\tenrm {I\kern-.18em{N}}\kern-.18em{I}}}}
\def\rational{{\hbox{\tenrm {{Q\kern-.54em{\vrule height.61em width0.05em
depth-.04em}\hskip.54em}\kern-.34em{\vrule height.59em width0.05em
depth-.04em}}
\hskip.34em}}}

\nopagenumbers
\titlepage
\baselineskip=14pt
Universit\`a di Roma, Preprint n. 855 \hfill hepth@xxx/9204009

\title{2+1 Dimensional Quantum Gravity as a Gaussian Fermionic System and
the 3D-Ising Model}
\author{ Giuseppe Bonacina}
\address{ Dipartimento di Fisica, Universit\`a di Milano,
I-20133 Milano, Italy}
\author { Maurizio Martellini\foot{Permanent address:
Dipartimento di Fisica, Universit\'a di Milano, I-20133 Milano, Italy}}
\address{Dipartimento di Fisica, Universit\`a di Roma "La Sapienza",
I-00185 Roma, Italy
\break
I.N.F.N., sezione di Pavia, Pavia, Italy}
\andauthor{ Mario Rasetti}
\address {\it Dipartimento di Fisica, Politecnico di Torino, I-10129 Torino,
Italy
\break
I.N.F.M., Unit\'a Torino Politecnico, Torino, Italy}
\vskip .6truein
\abstract
We show that 2+1-dimensional Euclidean quantum gravity is equivalent, under
some mild topological assumptions, to a Gaussian fermionic system. In
particular, for manifolds topologically equivalent to $\Sigma_g\times\RrR$
with $\Sigma_g$ a closed and oriented Riemann surface of genus $g$, the
corresponding 2+1-dimensional Euclidean quantum gravity
may be related to the 3D-lattice Ising model before its thermodynamic limit.

\vskip .4truein
\line{PACS $\#$ 04.60 ; 05.50 ; 02.40.P\hfill}
\endpage
\baselineskip=18pt
\pageno=1
\hrule height 0 pt
\chapter{Introduction}
A few years ago, Witten
\Ref\wittenuno{E. Witten, Nucl. Phys. {\bf 311B} (1988/89) 46.}
showed that 2+1-dimensional
quantum gravity in a first order dreibein formalism is
 exactly soluble at the classical and quantum levels. The key
point in Ref. \wittenuno\ is the observation that the dreibein
 $e_{\mu}^a$ and the spin connection $\omega_{\mu}^a\equiv
\varepsilon^{abc} \omega_{\mu}^{bc}$ form a gauge field of the group $ISO(2,1)$
($ISO(3)$) in Lorentzian (Euclidean)
signature. Thus, the Einstein-Hilbert action
$$\left\{
\eqalign{&I ={k \over 2} \int_{M^3} \varepsilon^{\mu \nu \rho} e_{\mu}^a
R_{\nu \rho}^a (\omega) \cr
&R_{\nu \rho}^a (\omega) \equiv \partial_{\nu} \omega_{\rho}^a-\partial_{\rho}
\omega_{\nu}^a+[\omega_{\nu},\omega_{\rho}]^a \cr}\right.\eqno (1.1) $$
becomes the non-Abelian Chern-Simons action on $M^3$ with
gauge group $G=ISO(2,1)$ or $ISO(3)$ depending on the signature of the
3D-manifold $M^3$. Here, and in the following, we shall
assume $M^3$ closed and oriented than otherwise stated.
However, in this context the meaning
of solvability is quite obscure, since in Witten's approach solvability
is ascribed to the fact that the Hilbert space is essentially the space of
half-densities on the moduli space of flat $SO(2,1)$ $(SO(3))$ connections on
$\Sigma_g$, where
$\Sigma_g$ is a spacelike surface of $M^3$, which is a closed Riemann
surface of genus g. Witten resorts to a canonical quantization scheme,
which requires that $M^3$ is topologically $\sim \Sigma_g \times \RrR$. This
result clearly doesn't tell us anything on
the full quantum dynamics (\ie\  the inclusion in
the scheme of correlation functions), it is restricted to three manifolds
topologically equivalent to $\Sigma_g \times \RrR$ and it essentially
prescribes solving the
Hamiltonian constraints of 3D-QG before quantizing, which is a procedure
not necessarily equivalent to the standard (covariant) BRST-quantization.

In this note, we shall show how for each fixed generic (closed) three
manifold $M^3$, the partition function
$Z_{EQG}(M^3)$ of the Euclidean continued 3D-QG is equivalent (up to
a normalization factor) to the partition function of a Gaussian discrete
fermionic system whose action encodes the topological nature of $M^3$.
Namely, we shall represent $M^3$  as the manifold obtained by Dehn
surgery on $S^3$ along a link $L\subset S^3$
\Ref\rolfsen{
See \eg\  D. Rolfsen,{\it Knots and Links}, Publish or Perish, Inc.,
 (Washington 1976).}
and show that
(under some suitable conditions) $Z_{EQG}(M^3)=Z_{EQG}(L;S^3)$ is the partition
function of the free fermions propagating on the link diagram $D_L$ and on
its r-parallel versions.
One may then study
the unknown correlation functions of 3D-Euclidean quantum gravity,
in a way similar to the 3D-Ising model\rlap,
\Ref\kavalov{A. R. Kavalov and A. G. Sedrakyan,
Nucl. Phys. {\bf 285B} [FS19] (1987) 264 and references therein.}
 directly
in the fermionic formulation, which is Gaussian, rather than in the
(non-linear) Chern-Simons gauge description
\foot{Till now it is not known which
kind of generalized Jones polynomials give the non-Abelian Chern-Simons
theory with a non-compact gauge group $ISO(3)$ and on  a generic three
manifold $M^3$ not homeomorphic to $S^3$.}.
In particular, in the case $M^3$ is a hyperbolic three-manifold $N^3$,
$\partial N^3\not= \emptyset$,
we shall show that
$Z_{EQG}(N^3)$ is related to the reduced partition function
of the 3D-Ising model on the lattice $\Lambda$ which is embeddable in
$\partial N^3$
\foot{Clearly, the 3D-Ising model is understood before the thermodynamic
limit $N\rightarrow\infty$ is taken. In this limit, one should also
perform $g\rightarrow\infty$. In our picture, this amounts to considering
\Ref\bona{
G. Bonacina, M. Martellini and M. Rasetti, work in progress.}
a sort of double scaling limit (DSL)
\Ref\gross{
D. J. Gross and A. A. Migdal, Phys. Rev. Lett. {\bf 64} (1990) 127;
\nextline
Nucl. Phys. {\bf 340B} (1990) 333;
\nextline
E. Br\'ezin and V. Kazakov, Phys. Lett. {\bf 236B} (1990) 144;
\nextline
M. Douglas and S. Shenker, Nucl. Phys. {\bf 235B} (1990) 635.}
at the level of the reduced EQG-partition function on the Riemann surface
$\Sigma_g\sim\partial N^3$. Indeed, the
DSL is the usual way of formulating the genus expansion, \ie\ the sum over
all genera.}.
\chapter{The Euclidean 3D-QG Partition Function and the Alexander-Conway
Polynomial}
Our starting point is Witten's result about the partition function of
Euclidean 3D-QG. He shows
\Ref\wittendue{E. Witten,
Nucl. Phys. {\bf 323B} (1989) 113}
that if one selects a non-degenerate metric ${\bar g_{\alpha\beta}}$
on $M^3$ and a (background) flat $SO(3)$ spin connection
${\bar \omega_{\mu(\alpha)}^a}$, where $\alpha$
is a labelling index, and uses the Landau background gauge
condition
$${\bar D^{\mu}_{(\alpha)}}e_{\mu}^a={\bar D^{\nu}_{(\alpha)}}\mu_{\mu}^a=0
\eqno (2.1)$$
where ${\bar D^{\mu}_{(\alpha)}}\equiv {\bar g^{\mu\nu}}
{\bar D_{\nu(\alpha)}}$
is the covariant derivative with respect to the Levi-Civita
connection ${\bar\nabla_{\nu}}$ associated to
${\bar g_{\mu\nu}}$ plus the flat connection
${\bar \omega_{\mu(\alpha)}}$ of interest, \ie\
$\bar D_{\nu(\alpha)}=\bar \nabla_\nu+[\bar\omega_{\nu(\alpha)}, \cdot ]$,
then the Euclidean partition function for the
3D-QG including the Fadeev-Popov ghosts reads
$$\left\{
\eqalign{&Z_{EQG}(M^3) =\sum_{(\alpha)} Z_{EQG(\alpha)}(M^3) \cr
&Z_{EQG(\alpha)}(M^3)
={[\det'(\lapbar_{0(\alpha)})]^2 \over |\det'(\bar{\cal D}_{(\alpha)})|}=
{[\det'(\lapbar_{1(\alpha)})]^{1 \over 2} \over \det'(\lapbar_{2(\alpha)})}
[\det'(\lapbar_{3(\alpha)})]
^{3 \over 2} \cr}\right. \eqno (2.2)$$
(since: $\det'(\lapbar_{k(\alpha)})=\det'(\lapbar_{(3-k)(\alpha)})$) where
$\bar{\cal D}_{(\alpha)} =\ast\dbar_{(\alpha)}+\dbar_{(\alpha)}\ast$
(here $\ast$ is the Hodge duality
operator) and
$\lapbar_{i(\alpha)}\equiv(\dbar_{\mu(\alpha)}\dbar^{\mu}_{(\alpha)})_i$
is the Laplacian operator acting
on twisted i-forms. Furthermore $\det'(\ast)$ in (2.2) is a functional
determinant, regularized, for instance, by zeta-function technique
\Ref\seeley{
R. Seeley, Proc. Symp. Pure Math. {\bf 10} (1966) 288.}
and omitting zero-modes. Equation (2.2) is derived under the following
assumptions:
\item{i)}  that  the moduli space ${\cal N}$ of flat SO(3) connections
modulo $SO(3)$-gauge transformations consists of
finitely many points, and $\bar\omega_{\mu(\alpha)}$
is an arbitrary representative
of ${\cal N}$. If $M^3\sim\Sigma_g \times\RrR$, is an orientable closed
Riemann surface of genus $g$, ${\cal N}$ has connected components corresponding
to Euler classes $2g-2$, $2g-3$,..., $-(2g-2)$. Here, the relevant component
is that, say $\bar{\cal N}\in {\cal N}$, of maximal Euler class $2g-2$
(Ref. \wittendue );
\item{ii)} that all ${\cal N}$ connections are irreducible.
Of course, these
conditions on ${\cal N}$ drastically restrict the allowed topologies of
$M^3$, however, as this
set of ``good topologies'' is not empty, it is reasonable to work out
a quantization scheme for Euclidean 3D-gravity only for this particular set of
topological three manifolds.
On the basis of such assumption, we notice first that
the ratio of determinants in (2.2) is in fact the Ray-Singer
analytic torsion
\Ref\ray{
D. B. Ray and I. Singer, Adv. Math. {\bf 7} (1971) 145;\nextline
Ann. Math. {\bf 98} (1973) 154.}
(R.S.-torsion), $T_{\rho(\alpha)}(M^3)$,
relative to the orthogonal representation
$\rho_{(\alpha)}:\pi_1(M^3)\rightarrow O(3)$ (\ie\ the i-forms on the universal
cover of $M^3$ transform according to $\rho_{(\alpha)}$).
This is a topological invariant
of $M^3$\foot{Therefore, it is independent of the metric used in the
gauge fixing and Fadeev-Popov terms.}, which labels homotopy equivalent
spaces.

Let us assume that $\rho_{(\alpha)}$ is acyclic, \ie\ that
$H^\ast(M^3,\rho_{(\alpha)})$ is zero.
Cheeger and M\"uller have shown
\Ref\cheeger{
J. Cheeger, Ann. Math. {\bf 109} (1979) 259;\nextline
W. M\"uller, Adv. Math. {\bf 28} (1978) 233}
that in this case the R.S.-torsion $T_{\rho(\alpha)}(M^3)$ is equivalent
to the so-called Reidemeister torsion
\Ref\milnor{
J. Milnor, Bull. Ann. Math. Soc. {\bf 72} (1966) 358}
(R.-torsion), $\tau_{\rho(\alpha)}(M^3)\in\RrR^+$, which is a non-homotopy
topological invariant
that may be computed from the twisted (by $\rho_{(\alpha)}$)
cochain complex associated to $M^3$
by the suitable alternating product of determinants. In this case,
therefore, one may set\foot{Fried
\Ref\fried{
D. Fried, Invent. Math. {\bf 84} (1986) 523;
\nextline
S. Della Pietra and V. Della Pietra, {\it Analytic Torsion and Finite Group
Actions}, IAS Preprint (1989).}
has shown that this identification survives also in the non acyclic case
if $\rho_{(\alpha)}$ is orthogonal.}:
$$Z_{EQG(\alpha)}(M^3)=T_{\rho(\alpha)}(M^3)=\tau_{\rho(\alpha)}(M^3).
\eqno (2.3)$$
A few comments are in order:
\nextline
i) In the definition of the Reidemeister torsion $\tau_{\rho(\alpha)}(M^n)$
one must start
with a PL-manifold; but every 3-manifold may be triangulated and hence the
PL-assumption is unnecessary.
\nextline
ii) The representation $\rho_{(\alpha)}:\pi_1(M^3)
\rightarrow O(m)$ extends to a unique
ring homomorphism from the integral group ring $\ZzZ(\pi_1(M^3))$ to the ring
of all real $m\times m$-matrices. Now the Reidemeister torsion
$\tau_{\rho(\alpha)}(M^3)$,
as defined in Ref. \milnor , is an element of the so-called Whitehead group
$\bar K_1M_m(\RrR)$.  It is known (Ref. \milnor) that
$\bar K_1M_m(\RrR)\simeq \bar K_1\RrR$, which, in terms of
the Reidemeister torsion, is equivalent to saying that the representation of
$\pi_1(M^3)$ is given by the ring homomorphism
$\varphi_{(\alpha)}:\pi_1(M^3)\rightarrow F_0$, where $F_0$ is the commutative
multiplicative group of a field, \eg\ the field of real numbers $\RrR$.
Thus, we have:
$$\left\{
\eqalign{&\tau_{\rho(\alpha)}(M^3)= \tau_{\varphi(\alpha)}(M^3) \cr
&\rho_{(\alpha)}:\pi_1(M^3)\rightarrow O(m) \cr
&\varphi_{(\alpha)}:\pi_1(M^3)\rightarrow F_0(\RrR) \cr}\right. .
\eqno (2.4)$$

Our next step will be connecting Eq. (2.4) to an appropriate Alexander
polynomial
\Ref\alex{
W. Alexander, Trans. A. M. S. {\bf 30} (1928) 275.}
$\triangle_L$. For this purpose, we need the general definition of Dehn
surgery on a
3-variety. Following Lickorish\rlap,
\Ref\lick{
W. B. R. Lickorish, Ann. of Math. {\bf 76} (1962) 531.}
we may always construct $M^3$ by Dehn surgery along a link
$L=K_1\cup \ldots\cup K_n$ in $S^3$ in the following way:
$$M^3=[S^3-(K^\circ_{f_1}\cup
\ldots\cup K^\circ_{f_n})]\cup_h (K_{f_1}\cup \ldots\cup
K_{f_n})\equiv (S^3-L_f)\cup_h L_f\eqno (2.5)$$
where $K^\circ_{f_i}$ is the interior of $K_{f_i}$ and
$K_{f_i}$ is the {\it preferred framing} $f_i$ of each component $K_i$ of
$L\subset S^3$, \ie\ the map
$K_i\rightarrow K_{f_i}\sim S^1\times D^2$ in which the longitude $\lambda_i$
is
oriented in the same way as $K_i$ and the meridian $\mu_i$ has linking
number $+1$ with $L_i$. In (2.5), $h$ is the union of homomorphisms
$h_i:\partial K_{f_i}\rightarrow \partial K_{f_i}\subset M^3$ defined by
$h_\ast(\mu_i)=[J_i]=a_i\lambda_i+b_i\mu_i$, where $b_i$ is
the linking number between $L_i$ and $J_i$, whereas $J_i$ is a
specified fixed simple
closed curve in each $\partial K_{f_i}$ (clearly $a_i,b_i \in \ZzZ$). Notice
that the homeomorphism type of $M^3$ does not depend on the choice of $h$.
Then we have the
\bigskip\noindent
{\bf Fundamental Theorem} (Ref. \lick): every closed oriented three-manifold
may be obtained by Dehn surgery on a link $L$ in $S^3$ with surgery
coefficients
$\displaystyle{r_i={b_i\over a_i}}$.
\bigskip
The key point in the proof of the fundamental theorem and in what we shall
show later about the 3D-Ising model (Sec. 3), is the noticing that $h$,
defined equivalently as $h:S^3-\{K^\circ_{f_i}\}\rightarrow M^3-\{K_{f_i}\}$,
may be characterized by {\it an element $\tau$ of the genus-$g$ mapping class
group} (extensively defined in Sec. 3). We do it in the following way.
Every closed, orientable three-manifold $M^3$ admits, apart from the
Dehn surgery representation quoted previously, a Heegaard decomposition:
$M^3=H_1\cup_\tau H_2$, $\partial H_1\sim -\partial H_2\sim\Sigma_g$,
$\tau:\partial H_2\rightarrow\partial H_1$, where $H_1$ and $H_2$ are
handlebodies of genus $g$, \ie\ roughly speaking, orientable three-manifolds,
with boundary an algebraic curve $\Sigma_g$, which are obtained by attaching
$g$ disjoint handles $D^2\times [-1,1]$ to 3-balls $B^3$.
Let us choose now a Heegaard
decomposition of the same genus $g$ for $S^3$ and $M^3$ given by:
$S^3=H_1\cup_f H_2$ and $M^3=H_1\cup_{f'} H'_2$, where $f:\partial H_2
\rightarrow \partial H_1$ and $f':\partial H'_2\rightarrow \partial H_1$
($H_2$ and $H'_2$ are handlebodies of the same genus $g$). Thus, the above
homeomorphism $h$ may be defined as $I_d\times l$, where $l$ is the
homeomorphism $l:H_2-\{K^\circ _{f_i}\}\rightarrow H'_2-\{K^\circ_{f_i}\}$.
Now, one could show (Ref. \lick) that $l$ can always be obtained as an
extension
of the mapping class group element $\tau=(f')^{-1}f:\partial H_2\rightarrow
\partial H'_2\sim\partial H_2$, \ie\ $\tau$ belongs
to the group of isotopy classes of
orientation preserving self-diffeomorphisms of the orientable, closed Riemann
surface $\Sigma_g\sim\partial H_2$. This latter statement follows from the fact
that the handlebodies $H'_2$ and $H_2$ are homeomorphic since they have
the same genus $g$ (Ref. 2).

As Milnor (Ref. \milnor) first noticed, there is a close connection between
the Alexander polynomial $\triangle_L(t_1,\ldots,t_n)$
and the Reidemeister torsion $\tau_\varphi(S^3-L_f)$. Let us remind briefly the
definition of Alexander polynomial for a link
$L=K_1\cup \ldots\cup K_n$ in $S^3$. If $V_L$ is the exterior of $L$, \ie\
$V_L=S^3-L_f$,
then the homology
group $H_1(V_L)$ is canonically isomorphic to a free Abelian multiplicative
group with $n$ free generators $(t_1,\ldots,t_n)$. The
generator $t_i$ corresponds
to the homology class of a meridian $\mu_i$ of the preferred framing
$f_i$ of $K_i$, $f_i:K_i\rightarrow K_{f_i}$.
Clearly, if $L$ is a knot, that is if
$n=1$, we simply write $t_1$ instead of $t$. The integral group ring
$\ZzZ [H_1(V)]$ is identified via this correspondence with the Laurent
polynomial ring $\ZzZ [t_1,t_1^{-1},\ldots,t_n,t_n^{-1}]$. The Alexander
polynomial $\triangle_L(t_1,\ldots,t_n)$ of the link $L\subset S^3$
is this Laurent polynomial in the variables $(t_1,\ldots,t_n)$
determined up to multiplication by polynomials of the form
$\pm t_1^{r_1}\cdots t_n^{r_n}$ with integral $r_1,\ldots,r_n$. To summarize,
the Alexander polynomial is a homology invariant computable from the
one-dimensional homology group of the exterior of the link with appropriate
twisted coefficients. Then the Milnor-Turaev theorem
\Ref\milnordue{J. Milnor, Ann. Math. {\bf 76} (1962) 137;
\nextline
V. G. Turaev, Uspekhi Mat.Nauk. {\bf 41} (1986) 97.}
states that:

\bigskip\noindent
{\bf Milnor-Turaev Theorem} (Ref. \milnordue ): the Alexander polynomial
$\triangle_L$ of a link in $S^3$ is equal (up to a standard factor) to
the Reidemeister torsion $\tau_\varphi$ of the exterior of the link, \ie\
$$\tau_\varphi(S^3-L_f)\simeq \triangle_L(t_1,\ldots,t_n) \eqno (2.6)$$
where $\varphi$ is the ring homomorphism: $\pi_1(S^3-L_f)\rightarrow
\ZzZ[t_1,t_1^{-1},\ldots,t_n,t_n^{-1}]$.
Indeed, $S^3-K_{f_i}$ is a space with the homology of a solid torus, \ie\
$H_{i=0,1}(S^3-K_{f_i})=\ZzZ $ (otherwise zero), and hence in the case when
$M^3=S^3-L_f$ we may identify the commutative \underbar{multiplicative}
group $F_0$
defined in (2.4) with the Laurent polynomial ring $\ZzZ(t_i,t_i^{-1})$.
In general, if we have a set of homomorphisms $\rho_{(\alpha)}:\pi_1(S^3-L_f)
\rightarrow F_{0(\alpha)}(\ZzZ)
\simeq \ZzZ(t_{i(\alpha)},t_{i(\alpha)}^{-1})$, the
correspondence (2.6) will take the form $\tau_{\varphi(\alpha)}(S^3-L_f)\simeq
\triangle_{L(\alpha)}(t_i)$, where $\triangle_{L(\alpha)}(t_i)\equiv
\triangle_L(t_{i(\alpha)})$. However, the sums (in $(\alpha)$) over
$\tau_{\varphi(\alpha)}$ will become products (in $(\alpha)$) over
$\triangle_{L(\alpha)}$ since $\triangle_{L(\alpha)}$ is by definition
an element of the \underbar{multiplicative} group, namely of
$\ZzZ[H_{1(\alpha)}(S^3-L_f)]$.

To summarize, we have shown that if $M^3$ is obtained by Dehn surgery along
a certain link $L$ (with $n$ components) in $S^3$ with the preferred framing
$f$, then the 3D-Euclidean quantum gravity partition function in the
background Landau gauge is given by (up to some irrelevant normalization
factors):
$$Z_{EQG}[M^3=(S^3-L_f)\cup_h L_f]=\sum_{(\alpha)}
\tau_{\varphi(\alpha)}[(S^3-L_f)\cup_h L_f],
\eqno (2.7)$$
where in particular
$$\sum_{(\alpha)}\tau_{\varphi(\alpha)}(S^3-L_f)\simeq
\prod_{(\alpha)}\triangle_{L(\alpha)}(t_i) \qquad i=1,\ldots,n. \eqno (2.8)$$
In the next section we shall show that the argument of the sum in
(2.7) may be rewritten as the vacuum
average of the {\it link operator} $L$ in terms of the partition
function provided
by the Alexander polynomial $\triangle_L$. This shall be a natural consequence
of the fact that {\it $\triangle_L$ can be represented by a free fermionic
Berezin-type path integral} and L by
non-local composite free fermion operators.

It is also worth noticing in particular that if $M^3$ is
{\it a-priori} a fixed hyperbolic
three-manifold $N^3$, then it is homeomorphic, by the ring
homomorphism $\varphi$, to the
exterior of a knot $K$, \eg\ to $S^3-K_f$ if and only if
$K$ is not a satellite knot and a torus knot\rlap.
\Ref\thurston{
W. P. Thurston, Bull. Am. Math. Soc. {\bf 6} (1982) 357.}
Thus in that case, equation (2.7) becomes formally
$$Z_{EQG}(N^3)=\sum_{(\alpha)}\tau_{\varphi(\alpha)}
[\delta(S^3-K_f)]\simeq \prod_{(\alpha)}\delta^\ast \triangle_{K(\alpha)},
\eqno (2.9)
$$
where $\delta$ is the homomorphism $S^3-K_f\rightarrow N^3$, and $\delta^\ast$
denotes the lift to $\tau_{\varphi(\alpha)}$
(and hence $\triangle_K$) of $\delta$, \ie\
formally: $\tau_{\varphi(\alpha)}
[\delta(\cdot)]=\delta^\ast\tau_{\varphi(\alpha)}(\cdot)$.
$\delta^\ast$ may be obtained from
a matrix representation of the mapping class
group ${\cal M}_g$ canonically associated with the Heegaard decompositions of
genus $g$ for $S^3$ and $N^3$ (Ref. \lick)
\foot{Let us
recall that under
(any) framing $K_i\in L$ becomes a solid torus $T_i\equiv K_{f_i}$, $\partial
T_i\not =\emptyset$, and that the set of all homomorphisms, up to isotopies,
of a
surface is defined as the mapping class group of that surface\rlap.
\Ref\rasetti{
M. Rasetti, {\sl The Mapping Class Group in Statistical Mechanics: a concise
review}, in {\it Symmetries in Science 111}, Edited by B.Gruber and F.Iachello,
Plenum Publ.Co., (1989) and references therein;
\nextline
B. Wajnryb, Israel J. Math. {\bf 45} (1983) 157.}
Thus, the proof that $\delta\in {\cal M}_g$ follows directly
from Thurston construction (Ref. \thurston) of hyperbolic three-manifolds
$N^3$ as $N^3=N^{'3}-(T'_1,\ldots,T'_r)$, where $N'^3$ is hyperbolic and
the $T_i$'s are disjoint solid
tori obtained by framing a suitable link $L'$ with r components $K_i$ and by
the so called Lickorish twist theorem (Ref. \lick). In fact one may choose
Heegaard decompositions of the \underbar{same genus} for $S^3$ and $N^3$, \ie\
$S^3=H_1\cup_g H_2$, $N'^3=H'_1\cup_{g'} H'_2$, where
$g:\partial H_2\rightarrow \partial H_1$ and $g':\partial H'_2\rightarrow
\partial H'_1$. Here we assume that $\partial N'^3=\emptyset$.
Since all handlebodies
of a given genus are homeomorphic, choose any homeomorphism
$h:H_1\rightarrow H'_1$. It follows, as a consequence of Lickorish twist
theorem, that the homeomorphism
$f\equiv (g')^{-1}hg:\partial H_2\rightarrow \partial H'_2$ belonging to the
genus-$g$ mapping class group ${\cal M}_g$
extends to a
homeomorphism $\bar f:H_2-(T_1,\ldots,T_r)\rightarrow H'_2-(T'_1,\ldots,T'_r)$.
This extends the chosen $h:H_1\rightarrow H'_1$ to a homeomorphism
$\delta\equiv (h,\bar f):S^3-(T_1,\ldots,T_r)=S^3-L_f\rightarrow N'^3-
(T'_1,\ldots,T'_r)=N'^3-L'_f=N^3$. So, $\delta:S^3-L_f\rightarrow N$ carries
an action of the mapping class group ${\cal M}_g$.}.

A recent result by Kohno \Ref\ko{
T. Kohno, {\it Topological Invariants of 3-Manifolds Using Representations
of Mapping Class Groups I}, Nagoya preprint {\bf 6} (1990).}
allows to define topological invariants $K(M^3)$ of closed
orientable three-manifolds $M^3$ using the representations
of ${\cal M}_g$ in such a way that $K(M^3)$ is an invariant under
the Heegaard decomposition.  As noticed discussing the fundamental theorem,
any closed oriented 3-d
manifold admits a Heegaard decomposition which,
via Likorish theorem$\,^{[13]}$,
is naturally in one-to-one correspondence with an element of the
mapping class group
of genus equal to the Heegaard genus.  Khono's construction provides a
projective linear representation of ${\cal M}_g$
$$
\Phi_{k} : {\cal M}_g \rightarrow  GL ( Z_{k} ( \Gamma )) / \sim \quad ,
\eqno (2.10)
$$
\noindent where $k$ is a positive integer labelling representations, and
$Z_{k}$ is  a finite dimensional complex vector space, each element of
which is in one-to-one correspondence (via Khono's
$k+1$ admissible weights) with the edges of the dual graph
$\Gamma$, which is a trivalent graph associated with the {\sl pants}
decomposition of the Heegaard surface $\Sigma_{g}$.  $\sim$ denotes
equivalence with respect to a cyclic group implying only phase factors.
$K(M^{3})$ is, up to a normalization factor, the {\it trace} of $\Phi$
meant as the $00$ entry of the matrix $\Phi_{k} (h)\, , \, h \in {\cal M}_{g}$
being the Heegaard glueing homeomorphism,
with respect to the basis of $Z_{k}$.

The problem we are faced with here, on the other hand,
is the {\it construction of the
representations of the genus $g$ mapping class group ${\cal M}_g$ starting from
a Dehn surgery presentation for $M^3$}, for the following two reasons:
\item{i)}  the euclidean 3-d quantum gravity partition function (2.7) is a
topological invariant by way of the Reidemeister torsion of the 3-manifold
$M^3$, given by a Dehn surgery presentation, and it is therefore interesting
to investigate the relation between such an invariant and Kohno's
$K(M^3)$;
\item{ii)} in view of the features of the 3-d Ising model, whose
partition function is based on the whole set of irreducible
representations of the
mapping class group of $\Sigma_{g}$ (in turn presented in terms of
Dehn twists), and of the expected equivalence between
the two partition functions, we have to express Dehn's surgery invariants
in terms of Heegaard invariants.

In other words, the problem is understanding the connection existing
between $K(M^3)$ and the topological invariants $I(M^3)$ of the
three-manifolds $M^3$ obtained
\Ref\resh{
N. Reshetikhin and V. Turaev, Invent. Math. {\bf 103} (1991) 547;
\nextline
W. B. R. Lickorish, {\it 3-Manifolds and the Temperly-Lieb Algebra},
UCLA preprint (1990).}
by performing Dehn surgery on a framed link.

We sketch here a procedure to obtain representations of the mapping class
group from the Dehn surgery prescription, due to Kohno.
\Ref\cono{
T. Kohno, private communication.}
The data is a set of trivalent graphs $\gamma_i$ (3-holed spheres) and a link
$L_0$ such that $L_0\cup\gamma_i$ is a link $L \in S^3$.
At this point one has two options:
either performing surgery on the link $L=(L_0,\gamma_i)$ in $S^3$ with a
choice of framing (\eg\ the {\it preferred framing} discussed below Eq. (2.5))
thus obtaining a three-manifold $M^3$ as shown above, or, equivalently,
regarding the trinions $\gamma_i$ as the complementary space of the so-called
{\it pants decomposition} of a Riemann surface $\Sigma_g$. In other words, the
$\gamma_i$'s with, say, $i=1,\ldots ,n$, characterize a Riemann surface of
genus $\displaystyle{g={1\over 2}(n+2)}$. In this second
case, the three-manifold $M^3$ is obtained by
glueing, with a homeomorphism $f$, the cylinder $\Sigma_g\times I$ with another
copy  of $\Sigma_g$ (the link $L_0$ is inside the cylinder). Then it turns
out that $f$, called the {\it cylinder map}, belongs to the mapping class group
${\cal M}_g$ of $\Sigma_g$.
\foot{This construction of ${\cal M}_g$ representations may be understood also
in terms of the so-called
{\it plat} representation  of a link.
\Ref\birmi{
J. S. Birman, {\it Braids, Links and Mapping Class Group}, Ann. Math. Studies
{\bf 82}, PUP (1975).}
Namely, if $L$ denotes the link carrying the Dehn surgery before the framing,
we may represent it by a $(2g+2)$-plat. Recall that a $(2g+2)$-plat
representation of the above link $L$ in $S^3$ is a triad $(S^3,\Sigma_0,L)$
where $(S^3,\Sigma_0)$ is a Heegaard splitting of genus zero of $S^3$ which
separates $S^3$ into 3-balls $B^{(1)}$ and $B^{(2)}$ so that $B^{(i)}\cap L$ is
a collection of $g+1$ unknotted and unlinked arcs with $\partial(B^{(i)}\cap
L)$
a set of $2g+2$ points on $\Sigma_0\equiv\partial B^{(i)}$ for $i=1,2$. The
topological type of the triad $(S^3,\Sigma_0,L)$ is fully described by a
Heegaard sewing map $\varphi$ which is required to preserve the $2g+2$ points
in $\partial(B^{(i)}\cap L)$; hence, up to isotopy, it is an element of the
mapping class group ${\cal M}_{0,2g+2}\sim B_{2g+2}$ ($B_n$ is Artin's
$n$-strings braid group) of the $(2g+2)$-punctured sphere. Here,
${\cal M}_{g,n}$ stands for the mapping class group for an $n$-punctured genus
$g$ Riemann surface $\Sigma_{g,n}$. Then, as it is well known (Ref. \birmi),
if $\sigma_i$ is the standard braid generator of $B_{2g+2}$ which interchanges
the $i$-th and the $(i+1)$-th points of $\partial(B^{(1)}\cap L)$, $\sigma_i$
lifts to the Dehn twist $\tau_{C_i}$ ($i=1,\ldots,2g+1$) along the
non-contractible circle $C_i$ decomposing
the Heegaard surface $\Sigma_g$ in a standard way. We may
visualize $\Sigma_g$ as the 2-fold covering of the sphere $\Sigma_0$
branched over $\partial(B^{(i)}\cap L)$, $i=1,2$. Thus, ${\cal M}_g$ is
{\it minimally} generated by a homomorphic image of $B_{2g+2}$ and one further
element (Ref. 16).}  Two comments are now in order:
\item{a)} $f$ provides a representation
of the Heegaard decomposition;
\item{b)}  whereas the process leading from  the Heegaard decomposition to the
Dehn surgery (and  to the mapping class group representations)
is one-to-one (naturally up to Heegaard
equivalence), the inverse construction leading from Dehn surgery to Heegard
decomposition (and once more to a representation of ${\cal M}_{g}$) is not
necessarily one-to-one.

In other words, the surgery link $L$ depends on $f$, whereas $f$
in general does not depend on $L$ alone.

We argue that the two invariants derived one within the Dehn surgery scheme,
the other in the Heegaard decomposition, should be related.
To begin with, we recall that Cappell-Lee-Miller
\Ref\capp{
S. E. Cappell, R. Lee and E. Y. Miller, {\it Invariants of 3-Manifolds from
Conformal Field Theory}, Courant Institute preprint (1990).}
have recently shown, in the frame of a conformal field theory
approach to problem of topological
invariants of a 3-manifold, that the above invariants $K(M^3)$ and $I(M^3)$ are
the same up to a phase factor. In our specific case, if we take $I(M^3)\equiv
Z_{EQG}(M^3)$, where $Z_{EQG}(M^3)$ is given by Eq. (2.7), and recall that
$Z_{EQG}(M^3)$ is also
equal to the Reidemeister torsion $\tau(M^3)$, Eq. (2.3), the equivalence with
Kohno's invariant $K(M^3)$ follows immediately from the fact that
the Reidemeister torsion can distinguish homotopy equivalent spaces
\Ref\cohen{
M. Cohen, {\it A Course in Simple Homotopy Theory}, Springer-Verlag
(New York, 1973).}
just like Kohno's invariant $K(M^3)$.
For example, the invariant $K(M^3)$ can
distinguish the Lens spaces $L(7,1)$ and $L(7,2)$, which are
not homeomorphic three-manifolds with the same homotopy,
like $\tau(M^3)$
\Ref\dubro{
B.A. Dubrovin, A.T. Fomenko, and S.P. Novikov,
{\it Modern Geometry -- Methods and Applications}, Springer-Verlag
(New York, 1990).}.
Of course, the equivalence between $K(M^3)$ and $Z_{EQG}(M^3) \equiv
\tau (M^3)$, is up to a suitable irrelevant phase factor which,
in the physical picture of $Z_{EQG}(M^3)$ as a
path integral (see next section), can be always reabsorbed into the
functional measure.

Furthermore, the above problem can be seen in
the ampler framework of distinguishing
between homology and homotopy equivalence of manifolds in 3-d.
Two 3-manifolds, say $M$ and $N$, are said to be {\it simple homotopy}
equivalent if their $CW$ complexes can be obtained one from the other by
a $CW$ deformation (namely adding a finite sequence of cells).  The
question whether homotopy equivalence implies simple homotopy equivalence
was answered by Whitehead
\Ref\White{
J.H.C. Whitehead, Amer. J. Math. {\bf 72}, 1 (1952)}
by a theorem stating that the obstruction to such implication is just the
non-vanishing of the Witehead torsion, of which the Reidemeister
torsion is a representation.

As both the Kohno connection and our invariant (2.7) are simple homotopy
invariants, yet $\underline{\rm not}$ homotopy invariants, the above arguments,
together with the results of Turaev and Viro \Ref\tuvi{V.G. Turaev, and O.Y.
Viro, {\it State Sum Invariants of 3-manifolds and Quantum 6{\sl j} Symbols},
LOMI preprint E-4-91} who succeed in connecting the invariant constructed from
the $q$-$6\, j$ symbols of the quantum group $U_{q} (s\ell 2)$ (which has the
same semiclassical limit as our $EQG$ partition function) to the Kohno
invariant, we conjecture, and shall henceforth assume, the equivalence of
$I(M^3)$ and $K(M^3)$.

\chapter{Free Fermions and the 3D-Euclidean Quantum Gravity Partition Function}

The mapping class group ${\cal M}_g$ of an orientable 2-{\sl manifold}
$\Sigma_g$ of genus $g$ is defined as the group of path components ({\it i.e.}
modulo isotopy) of the group of all orientation preserving homeomorphisms of
$\Sigma_g$ . Baer-Nielsen's theorems gives us the equivalent definition : the
mapping class group of a surface is isomorphic to the outer automorphism group
of its fundamental group.

It is interesting to recall here a few basic facts about the representations of
${\cal M}_{g}$. First, one fixes the cut system ${\cal C}_0 \equiv \{ \alpha_1
, \dots , \alpha_g \}$, namely a collection of disjoint circles  on $\Sigma_g$
such that $\displaystyle{\Sigma_{g} \setminus \left [\, \bigcup_{i=1}^{g}
\alpha_{i}\right ]}$ is a connected manifold, isomorphic with a $2g$-punctured
sphere. The simplest choice  is $\alpha_{1}$ goes once around the first handle,
$\alpha_{i}\, , \, 2=1, \dots ,g$ goes once around the $g$-th handle separating
the $(i-1)$-th from the $i$-th hole. One defines then the new family of closed
simple curves on $\Sigma_g \, , \; \{ \omega_{i,j} \,;\, 1 \le i < j \le 2g \}
\,: \; \omega_{i,j}$ interlaces handles $i$ and $j$ [ more precisely,
$\omega_{i,j}$ enters hole $i$ , goes around handle $i$, comes out of hole
$(i-1)$, enters hole $j$, goes around handle $j$, comes out of hole $(j-1)$ and
closes]. Denote by ${\cal W}_{i,j}$ the Dehn's twist with respect to
$\omega_{i,j}$. One defines moreover the following new homeomorphisms of
$\Sigma_g \, : \, {\cal P} := {\cal A}_g {\cal B}_g {\cal A}_g$ , which is a
simple move permuting $\alpha_g$ and $\beta_g \, {\rm ;} \; {\cal L} := {\cal
B}_g {\cal A}_g {\cal A}_g {\cal B}_g$ , which reverses the orientation of
$\alpha_g \,$, and $\, {\cal T}_i := {\cal B}_{i} {\cal A}_{i} {\cal A}_{i+1}
{\cal B}_{i+1}\, , \, i = 1, \dots , g-1 \,$ which permutes the circles
$\alpha_i$ and $\alpha_{i+1}$.

The mapping class group
${\cal M}_g$ is {\sl generated} by $\; \{ {\cal L} \,; {\cal P} \,; {\cal A}_i
\,, i = 1, \dots , g \, ; {\cal T}_j , j = 1, \dots , g-1 ; {\cal W}_{i,j} \,,
1 \le i < j \le g \} \;$.

Let now ${\cal H}_0$ be the stabilizer
subgroup, generated by $\, \{ {\cal A}_i \, {\rm ;}
\, {\cal W}_{i,j} \} \,$, of elements of ${\cal M}_g$ which leave the circles
$\{ \alpha_i \}$ fixed ; and ${\cal H}$ the subgroup, generated by $\, \{
{\cal H}_0 \, {\rm ;} \, {\cal L} \, {\rm ;} \, {\cal T}_i \} \,$, of elements
which leave the cut system ${\cal C}_0$ invariant . ${\cal H}$ is defined by
the exact sequences :
\def\mapright#1{\smash{\mathop{\longrightarrow}\limits^{#1}}}
$$
1 \, \mapright{} \,{\cal H}_0 \, \mapright{} \,{\cal H} \, \mapright\vartheta
\pm S_g \, \mapright{} \, 1 \quad ;
$$
$$
1 \, \mapright{} \, [{\ZzZ} / 2{\ZzZ}]^g \, \mapright{} \pm S_g \mapright{}
S_g \, \mapright{} \, 1 \quad ;
$$
\noindent where $\vartheta ({\cal L}) \in [{\ZzZ} / 2{\ZzZ}]^g$ and $\vartheta
 ({\cal T}_i)$ is the transposition $(i,i+1)$ in the symmetric group $S_g$ .

All the relations of ${\cal M}_g$ are generated by $\{ {\cal H} ,\,{\cal P}
\}$:

\item{{\it (I)}} ${\cal P}$ commutes with ${\cal H}_g$ (the subgroup of
elements of  ${\cal H}$ which leave $\alpha_g$ and $\beta_g$ invariant) ;
\item{{\it (II)}} ${\cal P}^2 \equiv {\cal A}_g {\cal L} {\cal A}_g \, \in \,
{\cal H}$ ;
\item{{\it (III)}} ${\cal P} {\cal F} {\cal P} {\cal F} {\cal P} \, \in \,
{\cal H} \;$ whenever $\exists \;$:
\itemitem{\it (1).} $\,\,$ a circle $\gamma$ on
$\Sigma_g$
which intersects once transversally both $\alpha_g$ and $\beta_g$ and does not
intersect any other $\alpha_i \,, i \not= g \; ,$ and $\;$
\itemitem{\it (2).} $\,\,$ a
map ${\cal F} \in {\cal H}$ such that $\; [ {\cal P}{\cal F} ]^{-1} \, \gamma
\, {\cal P}{\cal F} = \beta_g \,$ ; $\, [ {\cal P}{\cal F} ]^{-1} \, \beta_g \,
{\cal P}{\cal F} = \alpha_g \,$ ; $\, [ {\cal P}{\cal F} ]^{-1} \, \alpha_g \,
{\cal P}{\cal F} = \gamma \;$.
\item{{\it (IV)}} ${\cal P}$ commutes with $\tilde {\cal F} {\cal P} {\tilde
{\cal F}}^{-1}$ where $\tilde {\cal F} \in {\cal H}$ maps the simple closed
curve $\tilde \beta$ encircling holes $(g-1)$ and $g$ onto $\beta_g$ .
\item{{\it (V)}} ${\cal P}{\cal F}_1 {\cal P}{\cal F}_2 {\cal P}{\cal F}_3
{\cal P}{\cal F}_4 {\cal P} \in {\cal H} \;$ whenever $\exists \; $:
\itemitem{\it (1).}
$\,\,$ a circle $\delta$ on $\Sigma_g$ which intersects once transversally both
$\alpha_{g-1}$ and $\beta_g$ and does not intersect $\beta_{g-1}$ nor any other
$\alpha_i , i \not= g-1 \; ,$ and $\;$
\itemitem{\it (2).} $\,\,$ the maps
${\cal F}_j \in
{\cal H} ; j = 1, \dots , 4 \,$ satisfy -- upon defining $\;
{\cal E}_{(0)} := \IiI\, ; \,{\cal E}_{(n)} := {\cal E}_{(n-1)}
{\cal P}{\cal F}_n \, ; n=1, \dots ,4 \;$  (in terms of which the
element of ${\cal H}$ we are considering reads ${\cal E}_{(4)} {\cal P}$) --
the four relations :
$\quad {\cal E}_{(n)} \beta_{g-1} {\cal E}_{(n)}^{-1} =
\beta_g \; ; n=1, \dots ,4$.  When $\Sigma_g$ has no
punctures the isotropy subgroup ${\cal H}$ is included in
the exact sequence
$$
{\ZzZ} \, \mapright{} \, {\ZzZ}^g \oplus B_{2g-1} \, \mapright{} {\cal H}
\, \mapright{} \pm S_g \, \mapright{} \, 1 \; ;
$$
\noindent where $B_{2g-1}$ is the {\sl Artin} coloured {\sl braid group} over
$(2g-1)$ strings, whereas $\pm S_g$, the group of
signed permutations of $g$ objects, isomorphic with the group of $g \times
g$ matrices having just one non-zero entry, equal to $\pm 1$, in each row and
column.

The above presentation allows deriving information about faithful
representations of ${\cal M}_g$
\Ref\montor{A. Montorsi and M. Rasetti, {\sl The mapping Class Group :
Homology and Linearity}, in {\it Group Theoretical Methods in Physics},
H. D. D\"obner and T. D. Palev, eds.; World Scientific Publ. Co.; Singapore,
1988.}
\Ref\ivan{N. V. Ivanov, Uspekhi Mat. Nauk {\bf 42}:3, 49 (1987).}
For $g=1$ ,
${\cal M}_g \sim SL(2,{\ZzZ})$, the {\sl classical} (as opposed to the
{\sl Teichm\"uller} or {\sl many-handled}) Modular Group. The
related moduli space  is a space whose points correspond to
conformal isomorphism classes of tori. For arbitrary $g > 1$, upon denoting
by ${\cal I}(\Sigma_g)$ the set of isotopy classes of all the closed (non
oriented) curves
enbedded in $\Sigma_g$ , and by $\Phi_g$ any foliation whose leaves
are geodesics for for some  metric on $\Sigma_g$ (since $\Sigma_g$ has negative
Euler characteristics, the metric is hyperbolic), with transverse measure
$\mu_{\perp}$, we have the following results. $\mu_{\perp}(\bullet )$,
which  is a positive real function
assigning to
each arc $\sigma \in \Sigma_g$ transverse to the leaves of $\Phi_g$ and with
extremal points in $\Sigma_g \setminus \Phi_g$ an invariant weight, is
determined by
the conditions :

\item{{\it (a).}} $\mu_{\perp}(\sigma) = \mu_{\perp}(\sigma ')$ if $\sigma$ is
homotopic to $\sigma '$ through arcs transverse to $\Phi_g$ and with endpoints
in $\Sigma_g \setminus \Phi_g \,$;
\item{{\it (b).}} if $\, \sigma = \bigcup_i \sigma_i \,$ ; with $\, \sigma_i
\cap \sigma_j \subset \partial\sigma_i \cap \partial\sigma_j \,$ ; then
$\, \mu_{\perp}(\sigma ) = \sum_i \mu_{\perp}(\sigma_i) \;$ ;
\item{{\it (c).}} $\mu_{\perp}(\sigma ) \not= 0 \,$ if $\, \sigma \cap \Phi_g
\not= \emptyset \; .$

The collection of all these measured geodesic foliations constitutes a space
$\Xi_g$ on which ${\cal M}_g$ acts in a natural way.  In particular, in
this (faithful) representation, the
elements $m \in {\cal M}_g$ are classified according to the following scheme :
$m$ is said to be
\settabs 9\columns
\+ {\sl periodic} && , if it is of finite order in ${\cal M}_g$ ; \cr
\+ {\sl reducible} && , if there is a point in ${\cal I}(\Sigma_g )$ which is
invariant with respect to \cr
\+ && the element $m$ itself ; \cr
\+ {\sl pseudo-Anosov} && , if $\exists$ mutually transverse geodesic
foliations $\Phi_g^{(s)} , \,\Phi_g^{(u)} \in \Xi_g$ ($s$ \cr
\+ && stands for
{\sl stable} , $u$ for {\sl unstable}) , such that $\, m(\Phi_g^{(s)}) =
{1 \over \varepsilon} \Phi_g^{(s)} \,$ \cr
\+ && and $m(\Phi_g^{(u)}) = \varepsilon
\Phi_g^{(u)} \,$ for some real $\varepsilon > 1$ . \cr

In order to derive a faithful representation from our finite presentation, one
should first prove that $\underline{\rm no}$
normal subgroup ${\cal N}_{{\cal M}_g}$ of ${\cal M}_g$
can have all of its elements $\not= \IiI$ which are pseudo-Anosov, because
only in this case one can identify an
homeomorphism  $m_o \in {\cal N}_{{\cal M}_g}$ fixing some $\iota \in
{\cal I}(\Sigma_g)$ and then proceed in the construction of an {\sl induced}
faithful representation of ${\cal M}_g$ as a group of matrices (possibly with
entries in a field of characteristics $\not= 0$ or of anticommuting variables)
{}.

For example, let $\pi$ be a path on $\Sigma_g$ which
crosses the curve $\alpha_i$
at a finite number $\ell$ of points $\{ p_1^{(i)} , \dots , p_{\ell}^{(i)}
\}$ .  When we act on $\Sigma_g$ with ${\cal A}_i$ , the effect on $\pi$  is
that it is broken at each point $p_k^{(i)}$ and a copy of $\alpha_i$ is
inserted
 at the discontinuity in such a way as to coalesce (also in orientation) with
the adjacent fragments of $\pi$. Resorting to the property that on any compact
surface such as $\Sigma_g$
there exists at least a pair of essential simple closed curves, say $\gamma \,
, \, \gamma '$ , which {\sl fill} the surface but such that one can find
another essential closed curve $\tilde \gamma '$ , disjoint from $\gamma '$ ,
such that $\gamma \cup \tilde \gamma '$ does {\sl not} fill the surface, one
can show (Ref. \montor) that $\, {\cal D}_{\gamma} {\cal D}_{\gamma '}^{-1} \,$
is isotopic to a pseudo-Anosov map
\Ref\flp{A. Fathi, F. Laudenbach and V. Poenaru, Ast\'erisque, {\bf 66}-
{\bf 67}, 33 (1979)}.
Then $\, \gamma '' \equiv {\cal D}_{\gamma}
{\cal D}_{\gamma '}^{-1} \circ \tilde \gamma ' \,$ is a curve disjoint from
any essential simple curve $\tilde \gamma$ having  no intersections  with
$\gamma \cup \tilde \gamma '$ . Thus there exists a map
$$
{\cal D}_{\tilde \gamma '}^{-1} {\cal D}_{\gamma} {\cal D}_{\gamma '}^{-1}
{\cal D}_{\tilde \gamma '} {\cal D}_{\gamma '} {\cal D}_{\gamma}^{-1} \equiv
{\cal D}_{\tilde \gamma '}^{-1} {\cal D}_{\gamma ''}
$$
\noindent which fixes $\tilde \gamma$ and hence is {\sl not} pseudo-Anosov.

Considering the action of ${\cal M}_g$ on the projective space $\Xi_g$ of
measured geodesic foliations , Dehn's twists should be treated as maps with
parabolic action, since they are locally conjugate to the element
$\pmatrix{1&1\cr 0&1\cr} \in PSL (2 , {\ZzZ}) \,$. Moreover, recalling the
presentation of the fundamental group,
$$
\pi_{1} ( \Sigma_{g} ) \sim < {\cal A}_{1} , {\cal B}_{1} , \dots ,
{\cal A}_{g} , {\cal B}_{g} \, | \, \prod_{i=1}^{g} \left [\, {\cal A}_{i} ,
{\cal B}_{i}\, \right ] >
$$
\noindent where $\displaystyle{\left \{ {\cal A}_{i} , {\cal B}_{i} \, | \,
1 = 1, \dots , g\right \}}$ are assumed as a canonical basis for the first
homology group  $H_{1}(\Sigma_{g})$,
and noticing that its elements which act
parabolically
on the hyperbolic projective space are only those which may be freely
homotoped into cusps and that just these elements are non-Anosov, all that
remains to be done is to check -- by using the presentation --
whether ${\cal M}_g$ has a geometrically finite subgroup ${\cal S}_{{\cal
M}_g}$
on which it acts by conjugation.  Then, {\sl unless} the normal closure in
$\pi_1 (\Sigma_g )$ of the elements of the action of ${\cal M}_g$ on
${\cal S}_{{\cal M}_g}$ excludes all the cusp generators,
not all of its elements $\not= \IiI$ are pseudo-Anosov.

It is worth pointing out that this conclusion holds for $g \ge 2$ ,
when ${\cal M}_g$ there is a set of elementary homeomorphisms
equivalent to {\sl global
braids}. The corresponding matrix representation , when it exists , is that
induced from the {\sl monodromy} representation
associated with the {\sl Lefschetz fibration}
\Ref\maha{R. Mandelbaum and J. R. Harper, Can. Math. Soc. Conf. Proc. {\bf 2},
35 (1982).} of $\Sigma_g$ .

The approach to the Ising model which so far appeared
to be the most promising for extension
to the $d = 3$ case is that referred to as the {\sl Pfaffian} (or {\sl dimer})
method, whose formulation holds -- to a certain extent -- for any
number $d$ of dimensions. We briefly review here the formulation of such a
method that was proposed in Ref. 21
\Ref\rraassee{M. Rasetti, {\sl Ising Model on Finitely Presented Groups}, in
{\it Group Theoretical Methods in Physics}, M. Serdaroglu and E. In\"on\"u
eds.; Springer Verlag, Lecture Notes in Physics {\bf 180}; Berlin, 1983
\nextline plus references therein and in
\nextline G. Jacucci and M. Rasetti, J. Phys. Chem. {\bf 91}, 4970 (1987).}
as a possible candidate to attack some three-dimensional cases. It holds when
$\Lambda$ is homogeneous under some finitely presented (not necessarily finite)
group $G$ , and consists of a number of steps :

\item{[a]} the {\sl decorated} lattice $\Lambda_{\delta}$ is derived from
$\Lambda$ following Fisher's scheme
\Ref\fish{M. E. Fisher, J. Math. Phys. {\bf 7}, 1776 (1966).};
\item{[b]} the positional degrees of freedom in $\Lambda_{\delta}$ are
relabelled in terms of a set of anticommuting {\sl Grassmann} variables
$\eta_{g_{\ell}}\,$, in one-to-one correspondence with the group elements
$g_{\ell}$ of $G$ ;
\item{[c]} the group $G$ is extended to the group $\tilde {\cal G}$ in such a
way that all the bond orientations of $\Lambda_{\delta}$ compatible with the
combinatorial constraints imposed by the global generalization of the {\sl
Kasteleyn}'s theorem \foot{See Ref. \rraassee\ for a complete discussion of
this delicate issue. Intuitively, what is done in this approach is that a
generalized Kasteleyn's theorem is obtained by first extending the planar case
to the $2^{2g}$ covering of one of the surfaces, say  $\Sigma_g$, in which the
lattice $\Lambda_{\delta}$ is embedded, and by recovering then the non planar
case by summing over all possible boundary conditions (spinwise) for the
polygon whereby by suitable sides identifications $\Sigma_g$ -- which can of
course be thought of as a Riemann surface -- is obtained.  Such sum over all
possible choices of the boundary conditions is included in natural way in the
configuration sum giving the model partition function. Successively, that sum
is shown to be equivalent to summing over all possible ways of embedding
$\Lambda_{\delta}$ in a surface of genus $g$, and the latter in turn to be
identical to a sum over all the images of $\Sigma_g$ with respect to the
mapping class group, i.e. essentially with respect to all PL diffeomorphisms
modulo isotopy of $\sigma_g$ itself. This identifies the partition function
with the zeta function for the (infinite dimensional) set of flows induced by
the diffeomorphisms. Finally, the same partition function is expressed as a
product of (theta regularized) determinants, closely reminiscent of
Dirichlet-type zeta functions. These, resorting to Fried's definition
\Ref\Fried1{D. Fried, {\it Counting Circles}, in {\sl Dynamical Systems},
Springer-Verlag Lect. Notes Math. {\bf 1342}, 196 (1988)} can then be treated
as the formal dynamical zeta function associated with a flow at zero value of
its indeterminate. It has been recently proven by Moscovici and Stanton
\Ref\Mos{H. Moscovici, and R.J. Stanton, Invent. Math. {\bf 105}, 185 (1991)}
that this zeta function coincides with the $R$-torsion, with coefficients in
any flat acyclic bundle, for $\Sigma_g$. There is a complete consistency
between such result, and the similarity pointed out by Milnor\Ref\milnortre{J.
Milnor, {\it Infinite Cyclic Coverings}, in {\sl Topology of Manifolds},
Prindle, Weber \& Schmidt, Boston, 1968} between the algebraic formalism of
$R$-torsion in topology and zeta functions in the sense of Weil in dynamical
systems. Also, the result perfectly bridges the approach to the Ising model of
\rraassee and the present approach to quantum gravity, with the identity
exhibited by Ray and Singer in the second of refs. \ray expressing the
holomorphic analog of $R$-torsion for surfaces of genus >1 in terms of
classical Selberg zeta functions, whence the conjecture of equality between
$R$-torsion and analytic torsion, subsequently proven by Cheeger and M\"uller
\cheeger was derived.} to a non-planar case ({\it i.e.} to one in which the
lattice $\Lambda$ cannot be enbedded into a surface of genus zero, but can --
yet preserving the lattice coordination -- be enbedded into one, say $\Sigma_g$
of genus $g \ge 0$) and only those  can be obtained as the invariant (under
$\tilde {\cal G}$) set of configurations of the graph $\Gamma$ covering
$\Lambda_{\delta}$ $\, 2^{2g}$ times. \item{[d]} The partition function of the
model on $\Lambda$ is then given by
$$
{\cal Z}(\Lambda ) = \prod_{\alpha =1}^d \left \{
\cosh (K_{\alpha})\right \}^{N_{\alpha}}  {\rm Pf}\, {\tilde \Im} \quad ;
\quad \prod_{\alpha = 1}^{d} N_{\alpha} = N \quad ;
\eqno\eq
$$
\item{} where $\displaystyle{K_{\alpha} = {{J_{\alpha}}\over{k_{B} T}}}$
is the coupling constant of the model in direction $\alpha$, and $N$ the
toal number of sites of $\Lambda$.
${\tilde \Im}$ is the incidence matrix of $\Lambda_{\delta}$ , extended
with respect to $\tilde {\cal G}$ and $\, {\rm Pf} \,$ denotes the Pfaffian
(for a skew-symmetric matrix such as $\tilde \Im$ , ${\rm Pf} \equiv \sqrt
{\rm det}\,$) .
\item{[e]} If both $G$ and $g$ are finite, then $\tilde {\cal G}$ is finite,
and recalling that the regular
representation ${\cal R}$ of a finite group $\tilde {\cal G}$ is the direct sum
of its irreducible representations , labelled by an index $j$ , each contained
as many times as its dimension $dim\, j$, (3.1) reduces in a natural way to :
$$
{\cal Z}(\Lambda ) = \prod_{\alpha =1}^d \left \{
\cosh (K_{\alpha})\right \}^{N_{\alpha}} \prod_{j^{(F)}} \Bigl(
{\rm det}\,{\cal R}
\bigl[ \tilde \Im ^{(j^{(F)})} \bigr] \Bigr) ^{{1 \over 2}\, dim\, j^{(F)}}
\quad ;
\eqno\eq
$$
\item{} where the extra-index $F$ refers to {\sl Fermionic} representations ,
as required by the generalized Kasteleyn's theorem, and $\Im ^{(j)}$ is a
matrix of rank $j$ .

Let us recall first that the group ${\cal G}$ is called the {\sl extension} of
the group $G$ by the group $\Pi$ {\sl if} : having $G$ presentation
$G \approx \,< \Xi\, \vert
\, \Omega >\,$ , where $\Xi$ denotes the set of {\sl generators} and $\Omega$
the set of {\sl relations} , and similarly having $\Pi$ presentation
$\Pi \approx \,< \Upsilon\, \vert \, \Theta >\,$  ; we have the exact
sequence
$$
1\, \mapright{}\, G \, \mapright\iota \, {\cal G} \mapright\pi \, \Pi \,
\mapright{} \, 1 \quad ;
\eqno\eq
$$
\noindent and -- upon denoting by $\varphi$ a mapping which is the inverse of
the inclusion $\iota$ ; $\varphi : \Pi \rightarrow {\cal G}\,$ with $\,
\pi \circ \varphi = {\bf 1}_{\Pi}\,$ ; and by $\Upsilon ^{(\varphi)} \sim
\varphi (\Upsilon )$ the restriction of the relations of $\Pi$ to ${\cal G}$ --
${\cal G}$ has presentation :
$$
{\cal G} \approx \,< \Xi \, \cup \, \Upsilon ^{(\varphi)} \, \vert \,
\Omega \, \cup \, \{ \varphi ^{-1}(\upsilon )\, \xi \, \varphi (\upsilon )
\lambda_{\upsilon}^{-1}(\xi ) \, : \, \xi \in \Xi \, ;\, \upsilon \in
\Upsilon \} \, \cup \qquad\qquad
$$
$$
\qquad\qquad \cup \, \{ {\cal W}_{\vartheta}(\xi ) \, \vartheta (\varphi
(\upsilon )) \, : \, \vartheta \in \Theta \} > \; .
\eqno\eq
$$
\noindent where $\lambda_{\varpi} : G \rightarrow G \,$ is the automorphism
of $G$ induced by the action of the element $\varpi \in \Pi$ on $G$ :
$g_{\ell} \mapsto \varphi ^{-1}
(\varpi ) \, g_{\ell} \, \varphi (\varpi )$ ; and ${\cal W}_{\vartheta} \,$  is
some suitable word (one is to be selected for each $\vartheta \in \Theta$)
bringing
each element of ${\cal G}$ into the form ${\cal W}(\xi ) \cdot \gamma (\varphi
(\upsilon ))$ for some $\gamma \in \iota ( G ) \,$ .

Of course, each automorphism $\lambda_{\varpi}$ can be altered by an inner
automorphism of $G$ with no essential effect. If we factor out the
group of inner automorphisms we obtain a new mapping
$\, \kappa : \Pi \rightarrow Out\, G = Aut\, G \, / \, Inn \,G\,$
which is a homomorphism and is
basic for the extension, in that equivalent extensions define the same
homomorphism.   The triple $\{ \Pi , G , \kappa \}$ is called an {\sl abstract
kernel}
\Ref\zie{H. Zieschang, {\it Finite Groups of Mapping Classes}; Springer
Verlag, Lecture Notes in Mathematics {\bf 875}; Berlin, 1981.},
and a group ${\cal G}$ togehter with the exact
sequence (3.3) is
called an extension with respect to the abstract kernel if for $\gamma \in
\pi ^{-1}(\varpi )\, , \, \varpi \in \Pi \,$ , the automorphism of $G$ defined
by $g_{\ell} \mapsto \, \iota ^{-1} [ \gamma ^{-1} \iota (g_{\ell}) \,\gamma ]$
belongs to the equivalence class of $\kappa (\varpi )$.

The cases of physical interest are those in which $G$ is a {\sl Fuchsian} group
and $\Sigma_g$ is a {\sl factor surface} of $G$.  The center $\aleph$ of $G$
can therefore be considered as a $\Pi$-{\sl module} with an operation in the
equivalence class of $\kappa (\varpi )\, , \, \varpi \in \Pi\,$ , if $\Pi$ is
identified with the fundamental group $\pi_1 (\Sigma_g )$ . Considering now
the family of {\sl cohomology} groups $H^n (\Pi ,G) \, , \, n \ge 1\,$ of
$\Pi$ with coefficients in $G\,$ ({\it i.e.} the cohomology groups of the
{\sl cochain complexes} defined by $\, \{ {\cal C}^n (\Pi ,G)\, , \, \partial
^n \}_{n \in {\ZzZ}}\,$ , where $\, \partial ^n : {\cal C}^n (\Pi ,G)
\rightarrow {\cal C}^{n+1} (\Pi ,G)\,$ is the {\sl boundary operator} and
${\cal C}^n$ is an $n$-dimensional cochain\foot{Recall that
${\cal C}^n (\Pi ,G) \, , \, n \ge 1\,$ is the group of all functions
$\, f : \Pi^n \equiv \overbrace{\Pi \times \cdots \times \Pi}^{n\; {\rm times}}
\rightarrow G\,$ such that $f (\varpi_1 , \dots , \varpi_n ) = 0\,$ if some
$\varpi_i \, , \, 1 \le i \le n\,$ equals ${\bf 1}$ .}), one notices that
${\cal C}^3 (\Pi ,\aleph )$ -- upon regarding ${\cal C}^n (\Pi ,\aleph )$ as
an abelian group whose operation we write multiplicatively -- is zero (one
says that there is a trivial {\sl obstruction}).  The
theorem of Zieschang (Ref. \zie ) states then the extension ${\cal G}$ of
the abstract kernel $\{ \Pi , G , \kappa \}$ exists, and that ${\cal G}$ is a
proper subgroup of the mapping class group ${\cal M}_g$ .

Thus the homeomorphism $Ext : G \rightarrow \tilde {\cal G}$ required in $[c]$
\foot{It should be kept in mind that maps and spaces are to be
thought of in the $PL$ ({\sl piecewise-linear}) category, namely all
morphisms referred to in present discussion should be meant in the
corresponding definition as given in Ref. 24
\Ref\saund{C. P. Rourke, and B. J. Sanderson, Ann. Math. {\bf 87}, 1, 256,
and 431 (1968).}.}acts
locally by attaching
a Kasteleyn's phase to the circuits on $\Sigma_g$ homotopic to zero, and
globally by an extension by the fundamental group, {\it i.e.} mapping
$\pi_1 (\Sigma_g )$ to ${\ZzZ}_2$.  On the other hand, all possible
surfaces in which $\Lambda_{\delta}$ can be enbedded are equivalent from the
combinatorial point of view, and we can restrict to one {\it e.g.} by fixing
a cut system on $\Sigma_g$.  Moreover, as stated above,
the relations of the mapping
class group all follow from relations supported in
certain subsurfaces of $\Sigma_g$
finite in number and of genus at most 2.  There follows (Ref. \rraassee )
that the most general choice for $\tilde {\cal G}$ is :
$$
\tilde {\cal G} = \Re \bigotimes_{wr} S_{2g} \; ;
\eqno\eq
$$
\noindent where $\otimes_{wr}$ denotes the {\sl wreath}-product
\Ref\kerb{A. Kerber, {\it Representations of Permutation Groups}; Springer
Verlag, Lecture Notes in Mathematics {\bf 240}; Berlin, 1971.},
whose elements
can be taken to be all $2g \times 2g$ permutation matrices in which the non-
zero elements have been replaced by elements of $\Re$ ; whereas $\Re =
{\cal M}_g\, /\, {\cal H}\,$, namely the subgroup of diffeomorphisms of
$\Sigma_g$ which preserve the isotopy class of a maximal, unordered, non
separating system of $g$ disjoint, smoothly enbedded cycles (non
contractible and non isotopic), {\it e.g.} just the cut system
$\{ \alpha_i \, ;\, i = 1, \dots , g \} \,$ .  $\Re$ is
then essentially generated by the elements representing homology exchange
between any pair of circles $( \alpha_i , \alpha_j ) \, ;\, i,j = 1, \dots ,
g\,$.

Eq's (3.2) and (3.5) allow us now to write the {\sl free energy} ${\cal F}
\equiv - \kappa_B T \ln {\cal Z}$ as
$$
- \beta {\cal F} =
\sum_{\alpha =1}^d N_{\alpha} \ln \cosh (\beta J_{\alpha}) \, + \,
{1\over 2} \sum_{j^{(F)}} dim\, j^{(F)} \, {\rm Tr}
\Bigl( \ln {\cal R} \bigl[ \tilde \Im ^{(j^{(F)})} \bigr] \Bigr) \; ;
\eqno\eq
$$
\noindent from which it appears clearly that while ${\cal Z}$ can be
expanded in terms of {\sl characters} of $\Re$, ${\cal F}$, as given by
the latter equation, could be rewritten in terms of invariant symmetric
functions for $\Re$.  The
coefficients of such an invariant expansion retain some of the original
combinatorial flavour of the problem : they count the numbers of {\sl words}
in $\Re$ equivalent to the identity, {\it i.e.} provide a solution for the
Dehn's word problem for the subgroup ${\cal H}$ of ${\cal M}_g$.

An {\it unexpected bridge} between $Z_{EQG}(N^{3})$ as given
in (2.9), and ${\cal Z}(\Lambda )$ as given in (3.1),
can be cast by the following
argument.  Recalling the representation of ${\cal Z}( \Lambda )$
as grassmannian path integral,
as given by Itzykson
\Ref\itzyk{C. Itzykson, Nucl. Phys. {\bf B 210} [{\bf SF 6}], 448, aND 477
(1982) and \nextline
S. Samuel, J. Math. Phys. {\bf 21}, 2806, 2815, and 2820 (1980).},
we consider for simplicity the particular case in
which $M^3$ is obtained by Dehn surgery along a knot K in $S^3$. It follows,
that the associated Eq. (2.7) is a special form of the generalized surgery
formula for a non-Abelian 3D-Euclidean Chern-Simons gauge theory defined over
a generic three-manifold $\tilde M^3$ suggested by Witten\rlap.
\Ref\wittentre{
E. Witten, Comm. Math. Phys. {\bf 121} (1989) 351.}
Witten, in Ref. \wittentre, argues that:
$$\left\{
\eqalign{&Z[\tilde M^3] =\sum_j h_0^j Z[M^3;R_j] \cr
&Z[M^3;R_j] \equiv \langle W_{R_j}(K)\rangle_{M^3} \cr}\right. , \eqno (3.7)$$
where $\tilde M^3=M^3\cup_h K_f$, $h$ is the glueing homeomorphism on the
solid torus $K_f$ and $Z[M^3;R_j]$ is the CS-partition function of $M^3$
with an extra Wilson line $W_{R_j}(K)$ in the $R_j$ representation (of the
CS-gauge group ${\cal G}$) included on the knot K. When the
CS-coupling
$k$ is an integer, using the techniques of rational conformal field theories
(see \eg\ Ref. \wittentre), one could show that $R_j$ is a finite-dimensional
modulus of the representation ring of ${\cal G}$
with $j<\infty$. Then it turns out
that the knot diagram $D_K$ parametrized by $R_j$ has a nice (equivalent)
interpretation
\Ref\murakami{
J. Murakami, {\it The Parallel Version of Link Invariants}, Osaka
Preprints (1987);
\nextline
H. R. Morton and P. M. Strickland, ``Jones Polynomials Invariants for Knots and
Satellites'', to appear on Math. Proc. Camb. Phil. Soc. (1991).}
in terms of the so-called ``r-parallel version'' $C\ast D_K$ of $D_K$. That is,
for any $j\in \{1,\ldots,N\}$ we associate a non-negative integer $C(j)$,
called the ``colouring'' of $D_K$, from the set $\{1,2,\ldots,n\}\in \ZzZ_+$.
Let $C(j)\ast D_K$ be the diagram which can be formed by taking $C(j)$-copies
all parallel , in the plane, to $D_K$. In this picture Eq. (3.7)
becomes:
$$Z[M^3;R_j]=\langle W_{R_j}(K)\rangle_{M^3}=
\langle W_{\cal R}[C(j)\ast D_K]\rangle_{M^3}\equiv \langle
C(j)\ast D_K\rangle_{M^3}, \eqno (3.8)$$
the symbol $W_{\cal R}$ denoting the Wilson line
in the fundamental representation
${\cal R}$ of ${\cal G}$. Similarly, one finds that the coefficients
$h_j\equiv h_0^j$ (in general complex numbers) can be written as
$h_j=h_{C^{-1}(\ZzZ_+)}\equiv \lambda_c$ by definition of the colouring map
$C$.
Therefore, one can also write Eq. (3.7) as (remind that $K$ denotes a knot):
$$Z[\tilde M^3]=\sum_{c\in C} \lambda_c \langle c\ast D_K\rangle_{M^3}.
\eqno (3.9)$$
Eq. (3.9) has recently been rigorously stated by Lickorish (Ref. \resh)
in the case of the one-variable Jones polynomial for $\langle c\ast D_K
\rangle_{M^3}$ if
$M^3=S^3$ and ${\cal G}=su(2)$.

It is now immediate to notice that the partition function (2.7)
of the 3D-Euclidean quantum gravity has the form (3.9) with
$\tilde M ^3=(S^3-K_f)\cup_h K_f$ and $M^3=S^3-K_f$, if one sets
$$
\langle\ldots\rangle_{(S^3-K_f)}\simeq\sum_{(\alpha)}
\tau_{\varphi(\alpha)}(S^3-K_f)\simeq\prod_{(\alpha)}\triangle_K(t_{(\alpha)})
\equiv\prod_{(\alpha)}\triangle_{K(\alpha)}(t), \eqno (3.10)$$
where we have used Eq. (2.8) and $\displaystyle{t_{(\alpha)}\in
H_{1(\alpha)}(S^3-K_f)\equiv{\rho_{(\alpha)}[\pi_1(S^3-K_f)] \over
[\ast ,\ast]}}$,
and one regards $C \ast D_K$
as an extra ``field'' on which to compute the vacuum-to-vacuum
expectation value given by the ``partition function'' $\prod_{(\alpha)}
\triangle_{K(\alpha)}$.
Now, such an identification of $\triangle_{K(\alpha)}$ with a certain
path integral for each $(\alpha)$ is just what one in fact has!

Indeed, Kauffman and Saleur
\Ref\kauffman{L. H. Kauffman and H. Saleur, Comm. Math. Phys. {\bf 141},
293 (1991)}
have recently shown that the Alexander-Conway polynomial of a knot $K$
is the fermionic path integral over free fermions propagating on
the knot diagram $D_K$. Their basic idea is to describe the tangle
diagram $D_K$ as a planar Feynman graph $\Gamma_K$
for a Gaussian fermionic theory. The Feynman graph is obtained
by projection of the tangle diagram on a two-dimensional planar four-valent
graph. To each crossing $i$ of an oriented tangle diagram one associates
four complex Grassmannian variables $\psi^\alpha_i,\ \psi_j^{\beta\dagger}$
where the labels $\alpha=\beta=$ up ($u$), down ($d$) refer to edges
going up and down with respect to the direction of the crossing
at the point $i$.

All $\psi$'s anticommute
$$[\psi_i^\alpha,\psi_j^\beta]_+=0;\ \alpha,\beta=n\ or\ d;\ i\not= j$$
and in particular $(\psi_i^\alpha)^2=0$. The Berezin path integral is defined
as usual by the rule
$$\int\Pi_i d\psi_i^u d\psi_i^{u\dagger} d\psi_i^d d\psi_i^{d\dagger}
\Pi_i\psi_i^u\psi_i^{u\dagger}\psi_i^d\psi_i^{d\dagger}=1.$$
At Lagrangian level, if along the link $(i,j)$ the edge is oriented from
vertex $i$ to vertex $j$, the propagator is
$\psi_i^{\alpha\dagger}\psi_j^\beta$ with labels $\alpha,\beta=u
\ or\ d$ depending on the particular configuration. For instance,
 the tangle
$D_K$ or equivalently the associated Feynman graph $\Gamma_K$, both
shown in \FIG\tangle{An example of the correspondence between
a knot diagram $D_K$ and its relative planar Feynman graph $\Gamma_K$ for
a Gaussian fermionic theory.}
Fig.~\tangle, correspond to the kinetic term
$\psi^{u\dagger}_i\psi^d_j$.
Thus, the Kauffman-Saleur's result is
\medskip
{\bf KS-Theorem} (Ref. \kauffman): the Alexander-Conway polynomial
$\nabla_K(q)$
\foot{The Alexander-Conway polynomial for a knot is defined by the skein
relation
\Ref\kauffmandue{
L. H. Kauffman, Topology {\bf 20} (1981) 101.}
$$\nabla_{K_+}(q)-\nabla_{K_-}(q)=q\nabla_{K_0}(q)$$
and by the normalization: $\nabla_K=0$ for $K=$ (unknot) and $\nabla_K=1$
for $K=$ (unknotted strand).}
for a fixed knot $K$ has the fermionic path integral representation
\foot{
As it is well known, the Berezin path integral in (3.11) gives the square of
the Pfaffian $Pf(M)\equiv \sqrt{\det[M(K;q)]}$.}
$$\left\{
\eqalign{
&\nabla_K(q) =\langle\psi_+|\psi_-\rangle\equiv N(K;q)
\int d\psi^\dagger d\psi\exp\left\{\sum_{\scriptstyle i,j\atop\scriptstyle
\alpha,\beta=u,d}\psi_i^{\alpha\dagger}M_{\scriptstyle\alpha,\beta\atop
\scriptstyle i,j}(K;q)\psi_j^\beta\right\} \cr
&N(K;q) \equiv q^{-L(K)-I(K)} \cr}\right. , \eqno (3.11)$$
where $M=[M_{\scriptstyle\alpha,\beta\atop \scriptstyle i,j}(K;q)]$, which
depends on the type of knot $K$ selected, is a $[2\times \#\
(crossings)]^{\otimes 2}$-matrix whose entries are $\pm 1$ or certain rational
functions of $q$. Furthermore, $N$ is a normalization factor specified by the
number of internal edges $I(K)$ (loops $L(K)$) of the Feynman graph $\Gamma_K$
associated to $D_K$.
\bigskip
Since the usual Alexander polynomial $\triangle_K(t)$ is given by
$\nabla_K(q)$ in terms of the formula (Ref. \kauffmandue)
$$\triangle_K(t)=\nabla_K(q\equiv \sqrt{t}-{1 \over \sqrt{t}}), \eqno (3.12)$$
it follows that the Gaussian Berezin path integral (3.11) extends also to
$\triangle_K(t)$ and hence to $\tau_\varphi(S^3-K_f)$ via Eq. (2.6). In
our case we have a family of Alexander polynomials $\triangle_{K(\alpha)}$,
thus we shall have $\triangle_{K(\alpha)}\equiv\triangle_K(t_{(\alpha)})=
\langle\psi_+|\psi_-\rangle_{(\alpha)}\propto\sqrt{Pf(M_{K(\alpha)})}$, where
$M_{K(\alpha)}$ inherits the dependence on the labelling $(\alpha)$ by
$t_{(\alpha)}$.

Collecting all together, when $M^3$ is obtained by Dehn surgery along a knot
$K$ in $S^3$ we have the formula:
$$Z_{EQG}[M^3=(S^3-K_f)\cup_h K_f]=\sum_{c\in C}\lambda_c
\prod_{(\alpha)}
\langle\psi_+|[c\ast D_K]
(\psi_i^\alpha\psi_j^{\beta\dagger})|\psi_-\rangle_{(\alpha)},
\eqno (3.13)$$
where $h\ni h_j=h_{C^{-1}(\ZzZ_+)}\equiv \lambda_c$, $c\in \ZzZ_+$,
and $[c\ast D_K](\psi\psi^\dagger)$ denotes the operator associated to the
c-parallel version of $D_K$
in the Kauffman-Saleur fermionic representation. Clearly, following
Ref. \kauffman, we may identify $1\ast D_K=D_K$ with the action.  For
instance,
to the trefoil diagram $D_{\cal T}$ and to the associated Feynman
graph $\Gamma_{\cal T}$,
shown in \FIG\tref{The trefoil, its planar representation and relative
correspondences.} Fig.~\tref corresponds the matrix element
$\langle\psi_+|(\psi^\dagger M({\cal T};t)\psi)|\psi_-\rangle$ where (Ref. 18)
$$\eqalign{
\psi^\dagger M({\cal T};t)\psi=&\psi_1^{d\dagger} \psi^u_2+\psi_1^{u\dagger}
\psi^d_2+\psi_2^{d\dagger}\psi_3^u+\psi_2^{u\dagger}\psi^d_3+\psi_3^{u\dagger}
\psi^d_1+ \cr
&( \sqrt{t}-{1\over \sqrt{t}})\sum_{i=1}^3(\psi_i^{u\dagger}\psi^u_i+
\psi^{d\dagger}_i\psi^d_i)+(t+{1\over t}-3)\sum_{i=1}^3\psi^{u\dagger}_i
\psi^d_i \cr}.$$
As we already observed, the c-parallel
version of $D_K$ is a link analogue of the technique used to characterize
the representation ring of a Lie algebra by tensor products of the
fundamental representation, rather than by the irreducibles. In  our fermionic
picture, this corresponds to using integer powers
$(\psi^\dagger M_K\psi)^{l(c)}$, $l(c)\in \ZzZ_+$, of
$\psi^\dagger M_K\psi$ to describe $c\ast D_K$. As a consequence of this
description, the quantity under the product symbol in
the r.h.s. of Eq. (3.13) becomes the Berezin path integral of a
polynomial in $\psi_i^{\alpha\dagger}\psi_j^\beta$ and it may be interpreted
as the Green's function obtained from
$\int d\psi^\dagger d\psi \exp(\psi^\dagger M_K\psi)$.\foot{Clearly, one
has that:
$$\langle\psi_+|(\psi^\dagger M_K\psi)^{l(c)}|\psi_-\rangle_{(\alpha)}\sim
({\partial \over \partial\beta})^{l(c)} \int d\psi^\dagger d\psi
\exp(\beta\psi^\dagger M_{K(\alpha)}\psi)|_{\beta=1}.$$}

It is not surprising at this point that there exist an intimate relationship
between torsion invariants and partition function which can be formally
interpreted as dynamical zeta functions and, on the one side, Gaussian
fermionic ({\sl i.e.} grassmannian) stochastic systems, on the other
topological invariants.  On one front we have representatives of anosov
flows entering the partition function, whose relative weight heavily exceeds
that of closed orbits of the discrete periodic set and hence enphasizes the
stochastic features of the model.  On the other side the same
partition function, which describes the global dynamical (and/or
thermodynamical) behaviour
of the system can be viewed as the generating
function of all closed loops ({\sl i.e.} links with possibly knotted
components) embedded in a Riemann surface $\Sigma_g$.
The problem of characterizing the asymptotic images of the manifold
${\cal M}_{{\cal X} \cap {\cal Y}}$
intersection between two sub-varieties ${\cal X}$ and ${\cal Y}$
of a given manifold ${\cal W}$ under iteration of the group of
diffeomorphisms of ${\cal W}$ as, say, ${\cal X}$ is kept fixed
($dim\, {cal W} = dim\, {\cal X} + dim\, {\cal Y}$,
$dim\, {\cal M}_{{\cal X} \cap {\cal Y}} = |\, dim\, {\cal X} - dim\,
{\cal Y}\, |$), has been recently studied by V.I. Arno'ld \Ref\arnold{
V.I. Arno'ld, private communication, to appear in the Proceedings of the
Stony Brooks Symposium in honour of J. Milnor, A. Phillips, ed.;
Publish or Perish Press, Boston}.  Under an ergodic hypothesis, well
motivated physically, one expects that the equilibrium features
of the model we are considering are indeed controlled just by these
images (where one identifies obviously ${\cal X}$ with $\Sigma_{g}$,
${\cal Y}$ with the isotopy-equivalents of $\Sigma_g$ itself, and
${\cal M}_{{\cal X} \cap {\cal Y}}$ with the set of loops generated by
intersection).  The result of Arno'ld implies that it is just the set of
topological invariants of ${\cal M}_{{\cal X} \cap {\cal Y}}$
and it alone which
completely characterizes the asymptotic action of the group of
diffeomorphisms of ${\cal W}$ (that in our correspondence can be thought
of as the manifold of dimension $dim\, \Sigma_g + 1$ (in general non Euclidean
also in the Ising case, due to the choice of boundary conditions).

A final comment is in order. The procedure described in Sect. 3,
whereby we have essentially mapped the $3D$ Euclidean gravity to a free-fermion
system over a lattice is an homeomorphism between the two theories.
This is due to the conceptual passage through the Ising
model, which allows us in principle to reconstruct from the lattice model
the whole group of diffeomorphisms in $3D$, $Diff^{3}$.  The profound
meaning of such a
reconstruction can be understood in the following way:
the $3D$ Euclidean quantum gravity partition function is clearly
invariant with respect to $Diff^{3}$, as it essentially coincides with
the Reidemeister torsion which is diffeomorphically invariant by its
very construction.  On the other hand, at the
fermionic lattice model level one has (ref. \kauffman ) a {\sl hidden quantum
symmetry} $U_{q}[ s\ell (1,1) ]$, in other words, the link fermions
$\psi_{i}^{\alpha} \, , \, {\psi_{j}^{\beta}}^{\dagger}$ have
a non-trivial statistics whose {\sl dual} is the quantum group
$U_{q}[ s\ell (1,1) ]$.  The key notion here is then the following:
the quantum group symmetry $U_{q} [ {\it g} ]$ of a lattice model appears
on very short distances of the order of the lattice step; in the continuum
limit, it appears at a single point, valued in the group of
the Kac-Moody algebra $\hat g$ associated with the Lie algebra $g$ of some
finite-dimensional Lie group $G$.\foot{This observation first due to Alekseev,
Faddeev and Volkov \Ref\alek{
A. Alekseev, L. Faddeev, and A. Volkov,  {\it The unravelling of the
quantum group structure in the WZNW theory}, CERN preprint TH-5981/91}
in their study of the $WZNW$-model, applies here as well, with $\hat g =
\hat{s\ell (1,1)}$.} Thus we can equivalently affirm that our Gaussian
fermionic system has a gauge symmetry of type $G \approx SU(1,1)$, since
the corresponding Kac-Moody group acts just as a local gauge symmetry.
It is known (ref. \wittenuno )that the action in the continuum theory of
this local gauge $A_1$ symmetry on the first order fields $( e_{\mu} ,
\omega_{\mu} ) \equiv A_{\mu}$ describing the $3D$ gravitational field
$g_{\mu , \nu}$ is, on shell, equivalent to the to the action of
$Diff^{3}$ on $A_{\mu}$.  This leads us to interpreting the presence of
the group of diffeomorphisms in 3 dimensions in the continuum Einstein
gravity theory as the manifestation of the {\sl quantum} internal symmetry
of the underlying lattice model (3.11).

\chapter{Conclusions}
Summarizing, we have shown that 3D-Euclidean quantum gravity in
first order dreibein formalism and in the Landau gauge, when
quantized on a generic three-manifold obtained by Dehn surgery along a
knot $K$ (link $L$)
in $S^3$, is equivalent to a Gaussian fermionic theory propagating
on the c-parallel versions of the knot (link) diagram $D_K$ ($D_L$). In
particular we have shown in the Berezin path integral picture that the
3D-EQG partition function $Z(N^3)$ for a 3D-hyperbolic manifold $N^3$ is
equivalent (up to some irrelevant normalization factor) to that one
${\cal Z}(\Lambda)$ of a 3D-Ising model on a lattice $\Lambda$ embedded
in $\RrR^3$.

Let us conclude with two remarks:
\nextline
i) In the previous section we have shown that 3D-Euclidean quantum gravity
admits a free fermion representation as well as the 3D-Ising model
(Ref. 26)
and that these two models seem in mutual relation. Furthermore, we have proved
that for fixed tridimensional topologies the
3D-EQG partition function is given
by a suitable Alexander-Conway polynomial, which can explicitly be computed
by combinatorial or (Gaussian) path integral techniques.
So, thanks to the afore mentioned
equivalence, we also have a computable algorithm for solving the quantum
3D-Ising model before performing the thermodynamic limit.

Now, a related question is whether these are the only 3D-models which allow a
free fermion description, or rather it is a general property of (integrable)
models in 3D.
\nextline
ii) We would like to notice that an indirect hint of the possible quantum
connection between the 3D-Euclidean quantum gravity partition function and the
semiclassical limit of a polynomial link invariant may be found -- in view
mainly of the recent work by Turaev -- in the analysis performed by Ponzano and
Regge.
\Ref\ponz{ G. Ponzano and T. Regge, in {\it Spectroscopy and Group Theoretical
Methods in Physics}, North-Holland Publ.Co. (Amsterdam 1968).}
In Ref. \ponz\ it was argued that when $M^3$ is ``close" to $S^3$, the path
integral of the 3D-Euclidean quantum gravity in the simplicial approximation
known as the Regge calculus
\Ref\regge{
T. Regge, Nuovo Cimento {\bf 19} (1961) 551.}
is actually proportional to the semiclassical (large angular momentum) limit of
the standard $su(2)$ $6j$-symbol. We conjecture that the realization of the
Regge Ponzano program of understanding the Feynman summation of histories for
the lattice 3D euclidean Einstein-Hilbert action as a sort of state model
associated with the Racah coefficients, can be fully completed at quantum level
by our eq. (2.9).\foot{Very recently,\Ref\ogu{H. Ooguri, and N. Sasakura,
{\sl Discrete and continuum approaches to three-dimensional quantum gravity},
KUNS Preprint 1088/HE(TH) RIMS-788, 1991} \Ref\mizog{S. Mizoguchi, and
T. Tada, {\sl 3-dimensional gravity from the Turaev-Viro invariant},
YITP/U Preprint 91-43, 1991} Ooguri {\sl et al.} and Mizoguchi {\sl et al.}
argued that the Turaev-Viro model provides indeed a $q$-analogue lattice
regularization of the Ponzano-Regge model, where the cut-off is given
by $\displaystyle{{1\over{\ln q}}}$.  In their approach the equivalence with
the 3D Euclidean quantum gravity in the form (1.1) follows as continuum
limit of the Turaev-Viro piecewise-linear model, equivalent to the limit
$q \rightarrow 1$.}

Such a conjecture is based on the following facts:

\item{a)} it is known that the  standard $su(2)$ $6j$-symbol is the
''semiclassical" limit ($q\rightarrow 1$) of the quantum $6j$ symbol
$$\left\{
\matrix{
a & b & c \cr
d & e & f \cr
}\right\}_q\equiv [D]_q \eqno (4.1)$$
\noindent of the quantum group $U_q(sl(2;\RrR))$;
\item{b)} an intrinsic combinatorial approach is known which allows
to associate with the quantum $6j$ symbols of $U_q(sl(2;\RrR))$ the
two-variable
HOMFLY-polynomial $P_K(q,z)$ \Ref\turaev{
V. G. Turaev, Inventiones Math. {\bf 92} (1987) 527;
\nextline
A. N. Kirillov and N. Yu. Reshetikhin, {\it Representations of the $U_q(sl(2))$
Algebra, q-Orthogonal Polynomials and Invariants of Links}, LOMI
preprint E-9-88.};
\item{c)} the Alexander polynomial $\triangle_K(t)$ entering eq. (2.9), is
a particular case of $P_K(q,z)$ when $q=1$ \Ref\millet{
See for example W. B. R. Lickorish and K. C. Millet,
Topology {\bf 26} (1987) 107.} and $\displaystyle{z = \sqrt{t} -
{1\over{\sqrt{t}}}}$.

On the other hand, the possibility of constructing directly ''quantum"
invariants for a closed 3-manifold $M$ from the $q$-$6j$ symbols has been
recently stressed by Turaev and Viro in Ref. \tuvi\ , supporting once more our
conjecture that the Regge Ponzano idea might be extended to full quantum level.
Indeed we may recall that the Turaev-Viro 3-manifold invariant of $M$ is given
by
$$
|M|_{q} \equiv C^{\cal V} \sum_{\{ Co\ell \}}^{< \infty} \prod_{i \in {\cal E}}
(-1)^{i} [ 2i+1 ]_{q} \prod_{\cal T} \left [ D \right ]_{q}  \eqno{(4.2)}
$$
\noindent where $q$ is a complex root of unit of a certain degree $k \in
\ZzZ_{+}$, $k \geq 0$, $C$ is a constant, ${\cal V}$, ${\cal E}$, ${\cal T}$
denote respectively the numbers of vertices, edges, and the set of tetrahedra
of the simplicial complex ${\cal X}$, $[ n ]_{q}$ is the quantum dimension,
whereas $Co\ell$ is the map which
associates with the edges of ${\cal X}$ elements of the set $\{ Co\ell \}$
of {\sl colours}
$\displaystyle{\left \{ 0, {1\over 2}, 1, \dots , {{k}\over{2}} \right \}}$
related in standard way to the framing map introduced in (2.5).

Notice that the construction of the invariant is associated with a specific
triangulation of $M$; however the main theorem of Ref. \tuvi\ show just that
$|M|_{q} \in \CcC$ does not depend in fact on the choice of the triangulation
${\cal X}$, namely it is a {\sl bona fide} topological invariant.

Moreover, Turaev and Viro show that
$$
|M|_{q} = | I_{k}(M) |^{2} \quad , \eqno{(4.3)}
$$
\noindent where $I_{k}(M)$ is the Dehn surgery invariant for $M$ discussed in
Sect. 2, and $\displaystyle{q = \exp \left ( {{2 \pi i}\over{k+2}} \right )}$.

\ack
The authors gretefully acknowledge inspiring discussions with T. Kohno,
L. Kauffman, and T. Regge.

\par \penalty-400 \vskip\chapterskip
   \spacecheck\referenceminspace \immediate\closeout\referencewrite
   \referenceopenfalse
   \line{\fourteenrm\hfil References\hfil}\vskip\headskip
   \input referenc.txa
   
\endpage
\par \penalty-400 \vskip\chapterskip
   \spacecheck\referenceminspace \immediate\closeout\figurewrite
   \figureopenfalse
   \line{\fourteenrm\hfil FIGURE CAPTIONS\hfil}\vskip\headskip
   \input figures.txa

\end